\documentclass[twocolumn,tighten]{aastex61}
\usepackage[]{txfonts}
\usepackage{natbib, twoopt}
\usepackage{float}
\usepackage{graphicx}
\usepackage{epstopdf}
\usepackage{enumitem}
\usepackage[caption=false]{subfig}
\usepackage{color}\definecolor{AliceBlue}{rgb}{0.94,0.97,1.00}
\definecolor{AntiqueWhite1}{rgb}{1.00,0.94,0.86}
\definecolor{AntiqueWhite2}{rgb}{0.93,0.87,0.80}
\definecolor{AntiqueWhite3}{rgb}{0.80,0.75,0.69}
\definecolor{AntiqueWhite4}{rgb}{0.55,0.51,0.47}
\definecolor{AntiqueWhite}{rgb}{0.98,0.92,0.84}
\definecolor{BlanchedAlmond}{rgb}{1.00,0.92,0.80}
\definecolor{BlueViolet}{rgb}{0.54,0.17,0.89}
\definecolor{CadetBlue1}{rgb}{0.60,0.96,1.00}
\definecolor{CadetBlue2}{rgb}{0.56,0.90,0.93}
\definecolor{CadetBlue3}{rgb}{0.48,0.77,0.80}
\definecolor{CadetBlue4}{rgb}{0.33,0.53,0.55}
\definecolor{CadetBlue}{rgb}{0.37,0.62,0.63}
\definecolor{CornflowerBlue}{rgb}{0.39,0.58,0.93}
\definecolor{DarkBlue}{rgb}{0.00,0.00,0.55}
\definecolor{DarkCyan}{rgb}{0.00,0.55,0.55}
\definecolor{DarkGoldenrod1}{rgb}{1.00,0.73,0.06}
\definecolor{DarkGoldenrod2}{rgb}{0.93,0.68,0.05}
\definecolor{DarkGoldenrod3}{rgb}{0.80,0.58,0.05}
\definecolor{DarkGoldenrod4}{rgb}{0.55,0.40,0.03}
\definecolor{DarkGoldenrod}{rgb}{0.72,0.53,0.04}
\definecolor{DarkGray}{rgb}{0.66,0.66,0.66}
\definecolor{DarkGreen}{rgb}{0.00,0.39,0.00}
\definecolor{DarkGrey}{rgb}{0.66,0.66,0.66}
\definecolor{DarkKhaki}{rgb}{0.74,0.72,0.42}
\definecolor{DarkMagenta}{rgb}{0.55,0.00,0.55}
\definecolor{DarkOliveGreen1}{rgb}{0.79,1.00,0.44}
\definecolor{DarkOliveGreen2}{rgb}{0.74,0.93,0.41}
\definecolor{DarkOliveGreen3}{rgb}{0.64,0.80,0.35}
\definecolor{DarkOliveGreen4}{rgb}{0.43,0.55,0.24}
\definecolor{DarkOliveGreen}{rgb}{0.33,0.42,0.18}
\definecolor{DarkOrange1}{rgb}{1.00,0.50,0.00}
\definecolor{DarkOrange2}{rgb}{0.93,0.46,0.00}
\definecolor{DarkOrange3}{rgb}{0.80,0.40,0.00}
\definecolor{DarkOrange4}{rgb}{0.55,0.27,0.00}
\definecolor{DarkOrange}{rgb}{1.00,0.55,0.00}
\definecolor{DarkOrchid1}{rgb}{0.75,0.24,1.00}
\definecolor{DarkOrchid2}{rgb}{0.70,0.23,0.93}
\definecolor{DarkOrchid3}{rgb}{0.60,0.20,0.80}
\definecolor{DarkOrchid4}{rgb}{0.41,0.13,0.55}
\definecolor{DarkOrchid}{rgb}{0.60,0.20,0.80}
\definecolor{DarkRed}{rgb}{0.55,0.00,0.00}
\definecolor{DarkSalmon}{rgb}{0.91,0.59,0.48}
\definecolor{DarkSeaGreen1}{rgb}{0.76,1.00,0.76}
\definecolor{DarkSeaGreen2}{rgb}{0.71,0.93,0.71}
\definecolor{DarkSeaGreen3}{rgb}{0.61,0.80,0.61}
\definecolor{DarkSeaGreen4}{rgb}{0.41,0.55,0.41}
\definecolor{DarkSeaGreen}{rgb}{0.56,0.74,0.56}
\definecolor{DarkSlateBlue}{rgb}{0.28,0.24,0.55}
\definecolor{DarkSlateGray1}{rgb}{0.59,1.00,1.00}
\definecolor{DarkSlateGray2}{rgb}{0.55,0.93,0.93}
\definecolor{DarkSlateGray3}{rgb}{0.47,0.80,0.80}
\definecolor{DarkSlateGray4}{rgb}{0.32,0.55,0.55}
\definecolor{DarkSlateGray}{rgb}{0.18,0.31,0.31}
\definecolor{DarkSlateGrey}{rgb}{0.18,0.31,0.31}
\definecolor{DarkTurquoise}{rgb}{0.00,0.81,0.82}
\definecolor{DarkViolet}{rgb}{0.58,0.00,0.83}
\definecolor{DeepPink1}{rgb}{1.00,0.08,0.58}
\definecolor{DeepPink2}{rgb}{0.93,0.07,0.54}
\definecolor{DeepPink3}{rgb}{0.80,0.06,0.46}
\definecolor{DeepPink4}{rgb}{0.55,0.04,0.31}
\definecolor{DeepPink}{rgb}{1.00,0.08,0.58}
\definecolor{DeepSkyBlue1}{rgb}{0.00,0.75,1.00}
\definecolor{DeepSkyBlue2}{rgb}{0.00,0.70,0.93}
\definecolor{DeepSkyBlue3}{rgb}{0.00,0.60,0.80}
\definecolor{DeepSkyBlue4}{rgb}{0.00,0.41,0.55}
\definecolor{DeepSkyBlue}{rgb}{0.00,0.75,1.00}
\definecolor{DimGray}{rgb}{0.41,0.41,0.41}
\definecolor{DimGrey}{rgb}{0.41,0.41,0.41}
\definecolor{DodgerBlue1}{rgb}{0.12,0.56,1.00}
\definecolor{DodgerBlue2}{rgb}{0.11,0.53,0.93}
\definecolor{DodgerBlue3}{rgb}{0.09,0.45,0.80}
\definecolor{DodgerBlue4}{rgb}{0.06,0.31,0.55}
\definecolor{DodgerBlue}{rgb}{0.12,0.56,1.00}
\definecolor{FloralWhite}{rgb}{1.00,0.98,0.94}
\definecolor{ForestGreen}{rgb}{0.13,0.55,0.13}
\definecolor{GhostWhite}{rgb}{0.97,0.97,1.00}
\definecolor{GreenYellow}{rgb}{0.68,1.00,0.18}
\definecolor{HotPink1}{rgb}{1.00,0.43,0.71}
\definecolor{HotPink2}{rgb}{0.93,0.42,0.65}
\definecolor{HotPink3}{rgb}{0.80,0.38,0.56}
\definecolor{HotPink4}{rgb}{0.55,0.23,0.38}
\definecolor{HotPink}{rgb}{1.00,0.41,0.71}
\definecolor{IndianRed1}{rgb}{1.00,0.42,0.42}
\definecolor{IndianRed2}{rgb}{0.93,0.39,0.39}
\definecolor{IndianRed3}{rgb}{0.80,0.33,0.33}
\definecolor{IndianRed4}{rgb}{0.55,0.23,0.23}
\definecolor{IndianRed}{rgb}{0.80,0.36,0.36}
\definecolor{LavenderBlush1}{rgb}{1.00,0.94,0.96}
\definecolor{LavenderBlush2}{rgb}{0.93,0.88,0.90}
\definecolor{LavenderBlush3}{rgb}{0.80,0.76,0.77}
\definecolor{LavenderBlush4}{rgb}{0.55,0.51,0.53}
\definecolor{LavenderBlush}{rgb}{1.00,0.94,0.96}
\definecolor{LawnGreen}{rgb}{0.49,0.99,0.00}
\definecolor{LemonChiffon1}{rgb}{1.00,0.98,0.80}
\definecolor{LemonChiffon2}{rgb}{0.93,0.91,0.75}
\definecolor{LemonChiffon3}{rgb}{0.80,0.79,0.65}
\definecolor{LemonChiffon4}{rgb}{0.55,0.54,0.44}
\definecolor{LemonChiffon}{rgb}{1.00,0.98,0.80}
\definecolor{LightBlue1}{rgb}{0.75,0.94,1.00}
\definecolor{LightBlue2}{rgb}{0.70,0.87,0.93}
\definecolor{LightBlue3}{rgb}{0.60,0.75,0.80}
\definecolor{LightBlue4}{rgb}{0.41,0.51,0.55}
\definecolor{LightBlue}{rgb}{0.68,0.85,0.90}
\definecolor{LightCoral}{rgb}{0.94,0.50,0.50}
\definecolor{LightCyan1}{rgb}{0.88,1.00,1.00}
\definecolor{LightCyan2}{rgb}{0.82,0.93,0.93}
\definecolor{LightCyan3}{rgb}{0.71,0.80,0.80}
\definecolor{LightCyan4}{rgb}{0.48,0.55,0.55}
\definecolor{LightCyan}{rgb}{0.88,1.00,1.00}
\definecolor{LightGoldenrod1}{rgb}{1.00,0.93,0.55}
\definecolor{LightGoldenrod2}{rgb}{0.93,0.86,0.51}
\definecolor{LightGoldenrod3}{rgb}{0.80,0.75,0.44}
\definecolor{LightGoldenrod4}{rgb}{0.55,0.51,0.30}
\definecolor{LightGoldenrodYellow}{rgb}{0.98,0.98,0.82}
\definecolor{LightGoldenrod}{rgb}{0.93,0.87,0.51}
\definecolor{LightGray}{rgb}{0.83,0.83,0.83}
\definecolor{LightGreen}{rgb}{0.56,0.93,0.56}
\definecolor{LightGrey}{rgb}{0.83,0.83,0.83}
\definecolor{LightPink1}{rgb}{1.00,0.68,0.73}
\definecolor{LightPink2}{rgb}{0.93,0.64,0.68}
\definecolor{LightPink3}{rgb}{0.80,0.55,0.58}
\definecolor{LightPink4}{rgb}{0.55,0.37,0.40}
\definecolor{LightPink}{rgb}{1.00,0.71,0.76}
\definecolor{LightSalmon1}{rgb}{1.00,0.63,0.48}
\definecolor{LightSalmon2}{rgb}{0.93,0.58,0.45}
\definecolor{LightSalmon3}{rgb}{0.80,0.51,0.38}
\definecolor{LightSalmon4}{rgb}{0.55,0.34,0.26}
\definecolor{LightSalmon}{rgb}{1.00,0.63,0.48}
\definecolor{LightSeaGreen}{rgb}{0.13,0.70,0.67}
\definecolor{LightSkyBlue1}{rgb}{0.69,0.89,1.00}
\definecolor{LightSkyBlue2}{rgb}{0.64,0.83,0.93}
\definecolor{LightSkyBlue3}{rgb}{0.55,0.71,0.80}
\definecolor{LightSkyBlue4}{rgb}{0.38,0.48,0.55}
\definecolor{LightSkyBlue}{rgb}{0.53,0.81,0.98}
\definecolor{LightSlateBlue}{rgb}{0.52,0.44,1.00}
\definecolor{LightSlateGray}{rgb}{0.47,0.53,0.60}
\definecolor{LightSlateGrey}{rgb}{0.47,0.53,0.60}
\definecolor{LightSteelBlue1}{rgb}{0.79,0.88,1.00}
\definecolor{LightSteelBlue2}{rgb}{0.74,0.82,0.93}
\definecolor{LightSteelBlue3}{rgb}{0.64,0.71,0.80}
\definecolor{LightSteelBlue4}{rgb}{0.43,0.48,0.55}
\definecolor{LightSteelBlue}{rgb}{0.69,0.77,0.87}
\definecolor{LightYellow1}{rgb}{1.00,1.00,0.88}
\definecolor{LightYellow2}{rgb}{0.93,0.93,0.82}
\definecolor{LightYellow3}{rgb}{0.80,0.80,0.71}
\definecolor{LightYellow4}{rgb}{0.55,0.55,0.48}
\definecolor{LightYellow}{rgb}{1.00,1.00,0.88}
\definecolor{LimeGreen}{rgb}{0.20,0.80,0.20}
\definecolor{MediumAquamarine}{rgb}{0.40,0.80,0.67}
\definecolor{MediumBlue}{rgb}{0.00,0.00,0.80}
\definecolor{MediumOrchid1}{rgb}{0.88,0.40,1.00}
\definecolor{MediumOrchid2}{rgb}{0.82,0.37,0.93}
\definecolor{MediumOrchid3}{rgb}{0.71,0.32,0.80}
\definecolor{MediumOrchid4}{rgb}{0.48,0.22,0.55}
\definecolor{MediumOrchid}{rgb}{0.73,0.33,0.83}
\definecolor{MediumPurple1}{rgb}{0.67,0.51,1.00}
\definecolor{MediumPurple2}{rgb}{0.62,0.47,0.93}
\definecolor{MediumPurple3}{rgb}{0.54,0.41,0.80}
\definecolor{MediumPurple4}{rgb}{0.36,0.28,0.55}
\definecolor{MediumPurple}{rgb}{0.58,0.44,0.86}
\definecolor{MediumSeaGreen}{rgb}{0.24,0.70,0.44}
\definecolor{MediumSlateBlue}{rgb}{0.48,0.41,0.93}
\definecolor{MediumSpringGreen}{rgb}{0.00,0.98,0.60}
\definecolor{MediumTurquoise}{rgb}{0.28,0.82,0.80}
\definecolor{MediumVioletRed}{rgb}{0.78,0.08,0.52}
\definecolor{MidnightBlue}{rgb}{0.10,0.10,0.44}
\definecolor{MintCream}{rgb}{0.96,1.00,0.98}
\definecolor{MistyRose1}{rgb}{1.00,0.89,0.88}
\definecolor{MistyRose2}{rgb}{0.93,0.84,0.82}
\definecolor{MistyRose3}{rgb}{0.80,0.72,0.71}
\definecolor{MistyRose4}{rgb}{0.55,0.49,0.48}
\definecolor{MistyRose}{rgb}{1.00,0.89,0.88}
\definecolor{NavajoWhite1}{rgb}{1.00,0.87,0.68}
\definecolor{NavajoWhite2}{rgb}{0.93,0.81,0.63}
\definecolor{NavajoWhite3}{rgb}{0.80,0.70,0.55}
\definecolor{NavajoWhite4}{rgb}{0.55,0.47,0.37}
\definecolor{NavajoWhite}{rgb}{1.00,0.87,0.68}
\definecolor{NavyBlue}{rgb}{0.00,0.00,0.50}
\definecolor{OldLace}{rgb}{0.99,0.96,0.90}
\definecolor{OliveDrab1}{rgb}{0.75,1.00,0.24}
\definecolor{OliveDrab2}{rgb}{0.70,0.93,0.23}
\definecolor{OliveDrab3}{rgb}{0.60,0.80,0.20}
\definecolor{OliveDrab4}{rgb}{0.41,0.55,0.13}
\definecolor{OliveDrab}{rgb}{0.42,0.56,0.14}
\definecolor{OrangeRed1}{rgb}{1.00,0.27,0.00}
\definecolor{OrangeRed2}{rgb}{0.93,0.25,0.00}
\definecolor{OrangeRed3}{rgb}{0.80,0.22,0.00}
\definecolor{OrangeRed4}{rgb}{0.55,0.15,0.00}
\definecolor{OrangeRed}{rgb}{1.00,0.27,0.00}
\definecolor{PaleGoldenrod}{rgb}{0.93,0.91,0.67}
\definecolor{PaleGreen1}{rgb}{0.60,1.00,0.60}
\definecolor{PaleGreen2}{rgb}{0.56,0.93,0.56}
\definecolor{PaleGreen3}{rgb}{0.49,0.80,0.49}
\definecolor{PaleGreen4}{rgb}{0.33,0.55,0.33}
\definecolor{PaleGreen}{rgb}{0.60,0.98,0.60}
\definecolor{PaleTurquoise1}{rgb}{0.73,1.00,1.00}
\definecolor{PaleTurquoise2}{rgb}{0.68,0.93,0.93}
\definecolor{PaleTurquoise3}{rgb}{0.59,0.80,0.80}
\definecolor{PaleTurquoise4}{rgb}{0.40,0.55,0.55}
\definecolor{PaleTurquoise}{rgb}{0.69,0.93,0.93}
\definecolor{PaleVioletRed1}{rgb}{1.00,0.51,0.67}
\definecolor{PaleVioletRed2}{rgb}{0.93,0.47,0.62}
\definecolor{PaleVioletRed3}{rgb}{0.80,0.41,0.54}
\definecolor{PaleVioletRed4}{rgb}{0.55,0.28,0.36}
\definecolor{PaleVioletRed}{rgb}{0.86,0.44,0.58}
\definecolor{PapayaWhip}{rgb}{1.00,0.94,0.84}
\definecolor{PeachPuff1}{rgb}{1.00,0.85,0.73}
\definecolor{PeachPuff2}{rgb}{0.93,0.80,0.68}
\definecolor{PeachPuff3}{rgb}{0.80,0.69,0.58}
\definecolor{PeachPuff4}{rgb}{0.55,0.47,0.40}
\definecolor{PeachPuff}{rgb}{1.00,0.85,0.73}
\definecolor{PowderBlue}{rgb}{0.69,0.88,0.90}
\definecolor{RosyBrown1}{rgb}{1.00,0.76,0.76}
\definecolor{RosyBrown2}{rgb}{0.93,0.71,0.71}
\definecolor{RosyBrown3}{rgb}{0.80,0.61,0.61}
\definecolor{RosyBrown4}{rgb}{0.55,0.41,0.41}
\definecolor{RosyBrown}{rgb}{0.74,0.56,0.56}
\definecolor{RoyalBlue1}{rgb}{0.28,0.46,1.00}
\definecolor{RoyalBlue2}{rgb}{0.26,0.43,0.93}
\definecolor{RoyalBlue3}{rgb}{0.23,0.37,0.80}
\definecolor{RoyalBlue4}{rgb}{0.15,0.25,0.55}
\definecolor{RoyalBlue}{rgb}{0.25,0.41,0.88}
\definecolor{SaddleBrown}{rgb}{0.55,0.27,0.07}
\definecolor{SandyBrown}{rgb}{0.96,0.64,0.38}
\definecolor{SeaGreen1}{rgb}{0.33,1.00,0.62}
\definecolor{SeaGreen2}{rgb}{0.31,0.93,0.58}
\definecolor{SeaGreen3}{rgb}{0.26,0.80,0.50}
\definecolor{SeaGreen4}{rgb}{0.18,0.55,0.34}
\definecolor{SeaGreen}{rgb}{0.18,0.55,0.34}
\definecolor{SkyBlue1}{rgb}{0.53,0.81,1.00}
\definecolor{SkyBlue2}{rgb}{0.49,0.75,0.93}
\definecolor{SkyBlue3}{rgb}{0.42,0.65,0.80}
\definecolor{SkyBlue4}{rgb}{0.29,0.44,0.55}
\definecolor{SkyBlue}{rgb}{0.53,0.81,0.92}
\definecolor{SlateBlue1}{rgb}{0.51,0.44,1.00}
\definecolor{SlateBlue2}{rgb}{0.48,0.40,0.93}
\definecolor{SlateBlue3}{rgb}{0.41,0.35,0.80}
\definecolor{SlateBlue4}{rgb}{0.28,0.24,0.55}
\definecolor{SlateBlue}{rgb}{0.42,0.35,0.80}
\definecolor{SlateGray1}{rgb}{0.78,0.89,1.00}
\definecolor{SlateGray2}{rgb}{0.73,0.83,0.93}
\definecolor{SlateGray3}{rgb}{0.62,0.71,0.80}
\definecolor{SlateGray4}{rgb}{0.42,0.48,0.55}
\definecolor{SlateGray}{rgb}{0.44,0.50,0.56}
\definecolor{SlateGrey}{rgb}{0.44,0.50,0.56}
\definecolor{SpringGreen1}{rgb}{0.00,1.00,0.50}
\definecolor{SpringGreen2}{rgb}{0.00,0.93,0.46}
\definecolor{SpringGreen3}{rgb}{0.00,0.80,0.40}
\definecolor{SpringGreen4}{rgb}{0.00,0.55,0.27}
\definecolor{SpringGreen}{rgb}{0.00,1.00,0.50}
\definecolor{SteelBlue1}{rgb}{0.39,0.72,1.00}
\definecolor{SteelBlue2}{rgb}{0.36,0.67,0.93}
\definecolor{SteelBlue3}{rgb}{0.31,0.58,0.80}
\definecolor{SteelBlue4}{rgb}{0.21,0.39,0.55}
\definecolor{SteelBlue}{rgb}{0.27,0.51,0.71}
\definecolor{VioletRed1}{rgb}{1.00,0.24,0.59}
\definecolor{VioletRed2}{rgb}{0.93,0.23,0.55}
\definecolor{VioletRed3}{rgb}{0.80,0.20,0.47}
\definecolor{VioletRed4}{rgb}{0.55,0.13,0.32}
\definecolor{VioletRed}{rgb}{0.82,0.13,0.56}
\definecolor{WhiteSmoke}{rgb}{0.96,0.96,0.96}
\definecolor{YellowGreen}{rgb}{0.60,0.80,0.20}
\definecolor{aliceblue}{rgb}{0.94,0.97,1.00}
\definecolor{antiquewhite}{rgb}{0.98,0.92,0.84}
\definecolor{aquamarine1}{rgb}{0.50,1.00,0.83}
\definecolor{aquamarine2}{rgb}{0.46,0.93,0.78}
\definecolor{aquamarine3}{rgb}{0.40,0.80,0.67}
\definecolor{aquamarine4}{rgb}{0.27,0.55,0.45}
\definecolor{aquamarine}{rgb}{0.50,1.00,0.83}
\definecolor{azure1}{rgb}{0.94,1.00,1.00}
\definecolor{azure2}{rgb}{0.88,0.93,0.93}
\definecolor{azure3}{rgb}{0.76,0.80,0.80}
\definecolor{azure4}{rgb}{0.51,0.55,0.55}
\definecolor{azure}{rgb}{0.94,1.00,1.00}
\definecolor{beige}{rgb}{0.96,0.96,0.86}
\definecolor{bisque1}{rgb}{1.00,0.89,0.77}
\definecolor{bisque2}{rgb}{0.93,0.84,0.72}
\definecolor{bisque3}{rgb}{0.80,0.72,0.62}
\definecolor{bisque4}{rgb}{0.55,0.49,0.42}
\definecolor{bisque}{rgb}{1.00,0.89,0.77}
\definecolor{black}{rgb}{0.00,0.00,0.00}
\definecolor{blanchedalmond}{rgb}{1.00,0.92,0.80}
\definecolor{blue1}{rgb}{0.00,0.00,1.00}
\definecolor{blue2}{rgb}{0.00,0.00,0.93}
\definecolor{blue3}{rgb}{0.00,0.00,0.80}
\definecolor{blue4}{rgb}{0.00,0.00,0.55}
\definecolor{blueviolet}{rgb}{0.54,0.17,0.89}
\definecolor{blue}{rgb}{0.00,0.00,1.00}
\definecolor{brown1}{rgb}{1.00,0.25,0.25}
\definecolor{brown2}{rgb}{0.93,0.23,0.23}
\definecolor{brown3}{rgb}{0.80,0.20,0.20}
\definecolor{brown4}{rgb}{0.55,0.14,0.14}
\definecolor{brown}{rgb}{0.65,0.16,0.16}
\definecolor{burlywood1}{rgb}{1.00,0.83,0.61}
\definecolor{burlywood2}{rgb}{0.93,0.77,0.57}
\definecolor{burlywood3}{rgb}{0.80,0.67,0.49}
\definecolor{burlywood4}{rgb}{0.55,0.45,0.33}
\definecolor{burlywood}{rgb}{0.87,0.72,0.53}
\definecolor{cadetblue}{rgb}{0.37,0.62,0.63}
\definecolor{chartreuse1}{rgb}{0.50,1.00,0.00}
\definecolor{chartreuse2}{rgb}{0.46,0.93,0.00}
\definecolor{chartreuse3}{rgb}{0.40,0.80,0.00}
\definecolor{chartreuse4}{rgb}{0.27,0.55,0.00}
\definecolor{chartreuse}{rgb}{0.50,1.00,0.00}
\definecolor{chocolate1}{rgb}{1.00,0.50,0.14}
\definecolor{chocolate2}{rgb}{0.93,0.46,0.13}
\definecolor{chocolate3}{rgb}{0.80,0.40,0.11}
\definecolor{chocolate4}{rgb}{0.55,0.27,0.07}
\definecolor{chocolate}{rgb}{0.82,0.41,0.12}
\definecolor{coral1}{rgb}{1.00,0.45,0.34}
\definecolor{coral2}{rgb}{0.93,0.42,0.31}
\definecolor{coral3}{rgb}{0.80,0.36,0.27}
\definecolor{coral4}{rgb}{0.55,0.24,0.18}
\definecolor{coral}{rgb}{1.00,0.50,0.31}
\definecolor{cornflowerblue}{rgb}{0.39,0.58,0.93}
\definecolor{cornsilk1}{rgb}{1.00,0.97,0.86}
\definecolor{cornsilk2}{rgb}{0.93,0.91,0.80}
\definecolor{cornsilk3}{rgb}{0.80,0.78,0.69}
\definecolor{cornsilk4}{rgb}{0.55,0.53,0.47}
\definecolor{cornsilk}{rgb}{1.00,0.97,0.86}
\definecolor{cyan1}{rgb}{0.00,1.00,1.00}
\definecolor{cyan2}{rgb}{0.00,0.93,0.93}
\definecolor{cyan3}{rgb}{0.00,0.80,0.80}
\definecolor{cyan4}{rgb}{0.00,0.55,0.55}
\definecolor{cyan}{rgb}{0.00,1.00,1.00}
\definecolor{darkblue}{rgb}{0.00,0.00,0.55}
\definecolor{darkcyan}{rgb}{0.00,0.55,0.55}
\definecolor{darkgoldenrod}{rgb}{0.72,0.53,0.04}
\definecolor{darkgray}{rgb}{0.66,0.66,0.66}
\definecolor{darkgreen}{rgb}{0.00,0.39,0.00}
\definecolor{darkgrey}{rgb}{0.66,0.66,0.66}
\definecolor{darkkhaki}{rgb}{0.74,0.72,0.42}
\definecolor{darkmagenta}{rgb}{0.55,0.00,0.55}
\definecolor{darkolive}{rgb}{0.33,0.42,0.18}
\definecolor{darkorange}{rgb}{1.00,0.55,0.00}
\definecolor{darkorchid}{rgb}{0.60,0.20,0.80}
\definecolor{darkred}{rgb}{0.55,0.00,0.00}
\definecolor{darksalmon}{rgb}{0.91,0.59,0.48}
\definecolor{darksea}{rgb}{0.56,0.74,0.56}
\definecolor{darkslate}{rgb}{0.18,0.31,0.31}
\definecolor{darkslate}{rgb}{0.18,0.31,0.31}
\definecolor{darkslate}{rgb}{0.28,0.24,0.55}
\definecolor{darkturquoise}{rgb}{0.00,0.81,0.82}
\definecolor{darkviolet}{rgb}{0.58,0.00,0.83}
\definecolor{deeppink}{rgb}{1.00,0.08,0.58}
\definecolor{deepsky}{rgb}{0.00,0.75,1.00}
\definecolor{dimgray}{rgb}{0.41,0.41,0.41}
\definecolor{dimgrey}{rgb}{0.41,0.41,0.41}
\definecolor{dodgerblue}{rgb}{0.12,0.56,1.00}
\definecolor{firebrick1}{rgb}{1.00,0.19,0.19}
\definecolor{firebrick2}{rgb}{0.93,0.17,0.17}
\definecolor{firebrick3}{rgb}{0.80,0.15,0.15}
\definecolor{firebrick4}{rgb}{0.55,0.10,0.10}
\definecolor{firebrick}{rgb}{0.70,0.13,0.13}
\definecolor{floralwhite}{rgb}{1.00,0.98,0.94}
\definecolor{forestgreen}{rgb}{0.13,0.55,0.13}
\definecolor{gainsboro}{rgb}{0.86,0.86,0.86}
\definecolor{ghostwhite}{rgb}{0.97,0.97,1.00}
\definecolor{gold1}{rgb}{1.00,0.84,0.00}
\definecolor{gold2}{rgb}{0.93,0.79,0.00}
\definecolor{gold3}{rgb}{0.80,0.68,0.00}
\definecolor{gold4}{rgb}{0.55,0.46,0.00}
\definecolor{goldenrod1}{rgb}{1.00,0.76,0.15}
\definecolor{goldenrod2}{rgb}{0.93,0.71,0.13}
\definecolor{goldenrod3}{rgb}{0.80,0.61,0.11}
\definecolor{goldenrod4}{rgb}{0.55,0.41,0.08}
\definecolor{goldenrod}{rgb}{0.85,0.65,0.13}
\definecolor{gold}{rgb}{1.00,0.84,0.00}
\definecolor{gray0}{rgb}{0.00,0.00,0.00}
\definecolor{gray100}{rgb}{1.00,1.00,1.00}
\definecolor{gray10}{rgb}{0.10,0.10,0.10}
\definecolor{gray11}{rgb}{0.11,0.11,0.11}
\definecolor{gray12}{rgb}{0.12,0.12,0.12}
\definecolor{gray13}{rgb}{0.13,0.13,0.13}
\definecolor{gray14}{rgb}{0.14,0.14,0.14}
\definecolor{gray15}{rgb}{0.15,0.15,0.15}
\definecolor{gray16}{rgb}{0.16,0.16,0.16}
\definecolor{gray17}{rgb}{0.17,0.17,0.17}
\definecolor{gray18}{rgb}{0.18,0.18,0.18}
\definecolor{gray19}{rgb}{0.19,0.19,0.19}
\definecolor{gray1}{rgb}{0.01,0.01,0.01}
\definecolor{gray20}{rgb}{0.20,0.20,0.20}
\definecolor{gray21}{rgb}{0.21,0.21,0.21}
\definecolor{gray22}{rgb}{0.22,0.22,0.22}
\definecolor{gray23}{rgb}{0.23,0.23,0.23}
\definecolor{gray24}{rgb}{0.24,0.24,0.24}
\definecolor{gray25}{rgb}{0.25,0.25,0.25}
\definecolor{gray26}{rgb}{0.26,0.26,0.26}
\definecolor{gray27}{rgb}{0.27,0.27,0.27}
\definecolor{gray28}{rgb}{0.28,0.28,0.28}
\definecolor{gray29}{rgb}{0.29,0.29,0.29}
\definecolor{gray2}{rgb}{0.02,0.02,0.02}
\definecolor{gray30}{rgb}{0.30,0.30,0.30}
\definecolor{gray31}{rgb}{0.31,0.31,0.31}
\definecolor{gray32}{rgb}{0.32,0.32,0.32}
\definecolor{gray33}{rgb}{0.33,0.33,0.33}
\definecolor{gray34}{rgb}{0.34,0.34,0.34}
\definecolor{gray35}{rgb}{0.35,0.35,0.35}
\definecolor{gray36}{rgb}{0.36,0.36,0.36}
\definecolor{gray37}{rgb}{0.37,0.37,0.37}
\definecolor{gray38}{rgb}{0.38,0.38,0.38}
\definecolor{gray39}{rgb}{0.39,0.39,0.39}
\definecolor{gray3}{rgb}{0.03,0.03,0.03}
\definecolor{gray40}{rgb}{0.40,0.40,0.40}
\definecolor{gray41}{rgb}{0.41,0.41,0.41}
\definecolor{gray42}{rgb}{0.42,0.42,0.42}
\definecolor{gray43}{rgb}{0.43,0.43,0.43}
\definecolor{gray44}{rgb}{0.44,0.44,0.44}
\definecolor{gray45}{rgb}{0.45,0.45,0.45}
\definecolor{gray46}{rgb}{0.46,0.46,0.46}
\definecolor{gray47}{rgb}{0.47,0.47,0.47}
\definecolor{gray48}{rgb}{0.48,0.48,0.48}
\definecolor{gray49}{rgb}{0.49,0.49,0.49}
\definecolor{gray4}{rgb}{0.04,0.04,0.04}
\definecolor{gray50}{rgb}{0.50,0.50,0.50}
\definecolor{gray51}{rgb}{0.51,0.51,0.51}
\definecolor{gray52}{rgb}{0.52,0.52,0.52}
\definecolor{gray53}{rgb}{0.53,0.53,0.53}
\definecolor{gray54}{rgb}{0.54,0.54,0.54}
\definecolor{gray55}{rgb}{0.55,0.55,0.55}
\definecolor{gray56}{rgb}{0.56,0.56,0.56}
\definecolor{gray57}{rgb}{0.57,0.57,0.57}
\definecolor{gray58}{rgb}{0.58,0.58,0.58}
\definecolor{gray59}{rgb}{0.59,0.59,0.59}
\definecolor{gray5}{rgb}{0.05,0.05,0.05}
\definecolor{gray60}{rgb}{0.60,0.60,0.60}
\definecolor{gray61}{rgb}{0.61,0.61,0.61}
\definecolor{gray62}{rgb}{0.62,0.62,0.62}
\definecolor{gray63}{rgb}{0.63,0.63,0.63}
\definecolor{gray64}{rgb}{0.64,0.64,0.64}
\definecolor{gray65}{rgb}{0.65,0.65,0.65}
\definecolor{gray66}{rgb}{0.66,0.66,0.66}
\definecolor{gray67}{rgb}{0.67,0.67,0.67}
\definecolor{gray68}{rgb}{0.68,0.68,0.68}
\definecolor{gray69}{rgb}{0.69,0.69,0.69}
\definecolor{gray6}{rgb}{0.06,0.06,0.06}
\definecolor{gray70}{rgb}{0.70,0.70,0.70}
\definecolor{gray71}{rgb}{0.71,0.71,0.71}
\definecolor{gray72}{rgb}{0.72,0.72,0.72}
\definecolor{gray73}{rgb}{0.73,0.73,0.73}
\definecolor{gray74}{rgb}{0.74,0.74,0.74}
\definecolor{gray75}{rgb}{0.75,0.75,0.75}
\definecolor{gray76}{rgb}{0.76,0.76,0.76}
\definecolor{gray77}{rgb}{0.77,0.77,0.77}
\definecolor{gray78}{rgb}{0.78,0.78,0.78}
\definecolor{gray79}{rgb}{0.79,0.79,0.79}
\definecolor{gray7}{rgb}{0.07,0.07,0.07}
\definecolor{gray80}{rgb}{0.80,0.80,0.80}
\definecolor{gray81}{rgb}{0.81,0.81,0.81}
\definecolor{gray82}{rgb}{0.82,0.82,0.82}
\definecolor{gray83}{rgb}{0.83,0.83,0.83}
\definecolor{gray84}{rgb}{0.84,0.84,0.84}
\definecolor{gray85}{rgb}{0.85,0.85,0.85}
\definecolor{gray86}{rgb}{0.86,0.86,0.86}
\definecolor{gray87}{rgb}{0.87,0.87,0.87}
\definecolor{gray88}{rgb}{0.88,0.88,0.88}
\definecolor{gray89}{rgb}{0.89,0.89,0.89}
\definecolor{gray8}{rgb}{0.08,0.08,0.08}
\definecolor{gray90}{rgb}{0.90,0.90,0.90}
\definecolor{gray91}{rgb}{0.91,0.91,0.91}
\definecolor{gray92}{rgb}{0.92,0.92,0.92}
\definecolor{gray93}{rgb}{0.93,0.93,0.93}
\definecolor{gray94}{rgb}{0.94,0.94,0.94}
\definecolor{gray95}{rgb}{0.95,0.95,0.95}
\definecolor{gray96}{rgb}{0.96,0.96,0.96}
\definecolor{gray97}{rgb}{0.97,0.97,0.97}
\definecolor{gray98}{rgb}{0.98,0.98,0.98}
\definecolor{gray99}{rgb}{0.99,0.99,0.99}
\definecolor{gray9}{rgb}{0.09,0.09,0.09}
\definecolor{gray}{rgb}{0.75,0.75,0.75}
\definecolor{green1}{rgb}{0.00,1.00,0.00}
\definecolor{green2}{rgb}{0.00,0.93,0.00}
\definecolor{green3}{rgb}{0.00,0.80,0.00}
\definecolor{green4}{rgb}{0.00,0.55,0.00}
\definecolor{greenyellow}{rgb}{0.68,1.00,0.18}
\definecolor{green}{rgb}{0.00,1.00,0.00}
\definecolor{grey0}{rgb}{0.00,0.00,0.00}
\definecolor{grey100}{rgb}{1.00,1.00,1.00}
\definecolor{grey10}{rgb}{0.10,0.10,0.10}
\definecolor{grey11}{rgb}{0.11,0.11,0.11}
\definecolor{grey12}{rgb}{0.12,0.12,0.12}
\definecolor{grey13}{rgb}{0.13,0.13,0.13}
\definecolor{grey14}{rgb}{0.14,0.14,0.14}
\definecolor{grey15}{rgb}{0.15,0.15,0.15}
\definecolor{grey16}{rgb}{0.16,0.16,0.16}
\definecolor{grey17}{rgb}{0.17,0.17,0.17}
\definecolor{grey18}{rgb}{0.18,0.18,0.18}
\definecolor{grey19}{rgb}{0.19,0.19,0.19}
\definecolor{grey1}{rgb}{0.01,0.01,0.01}
\definecolor{grey20}{rgb}{0.20,0.20,0.20}
\definecolor{grey21}{rgb}{0.21,0.21,0.21}
\definecolor{grey22}{rgb}{0.22,0.22,0.22}
\definecolor{grey23}{rgb}{0.23,0.23,0.23}
\definecolor{grey24}{rgb}{0.24,0.24,0.24}
\definecolor{grey25}{rgb}{0.25,0.25,0.25}
\definecolor{grey26}{rgb}{0.26,0.26,0.26}
\definecolor{grey27}{rgb}{0.27,0.27,0.27}
\definecolor{grey28}{rgb}{0.28,0.28,0.28}
\definecolor{grey29}{rgb}{0.29,0.29,0.29}
\definecolor{grey2}{rgb}{0.02,0.02,0.02}
\definecolor{grey30}{rgb}{0.30,0.30,0.30}
\definecolor{grey31}{rgb}{0.31,0.31,0.31}
\definecolor{grey32}{rgb}{0.32,0.32,0.32}
\definecolor{grey33}{rgb}{0.33,0.33,0.33}
\definecolor{grey34}{rgb}{0.34,0.34,0.34}
\definecolor{grey35}{rgb}{0.35,0.35,0.35}
\definecolor{grey36}{rgb}{0.36,0.36,0.36}
\definecolor{grey37}{rgb}{0.37,0.37,0.37}
\definecolor{grey38}{rgb}{0.38,0.38,0.38}
\definecolor{grey39}{rgb}{0.39,0.39,0.39}
\definecolor{grey3}{rgb}{0.03,0.03,0.03}
\definecolor{grey40}{rgb}{0.40,0.40,0.40}
\definecolor{grey41}{rgb}{0.41,0.41,0.41}
\definecolor{grey42}{rgb}{0.42,0.42,0.42}
\definecolor{grey43}{rgb}{0.43,0.43,0.43}
\definecolor{grey44}{rgb}{0.44,0.44,0.44}
\definecolor{grey45}{rgb}{0.45,0.45,0.45}
\definecolor{grey46}{rgb}{0.46,0.46,0.46}
\definecolor{grey47}{rgb}{0.47,0.47,0.47}
\definecolor{grey48}{rgb}{0.48,0.48,0.48}
\definecolor{grey49}{rgb}{0.49,0.49,0.49}
\definecolor{grey4}{rgb}{0.04,0.04,0.04}
\definecolor{grey50}{rgb}{0.50,0.50,0.50}
\definecolor{grey51}{rgb}{0.51,0.51,0.51}
\definecolor{grey52}{rgb}{0.52,0.52,0.52}
\definecolor{grey53}{rgb}{0.53,0.53,0.53}
\definecolor{grey54}{rgb}{0.54,0.54,0.54}
\definecolor{grey55}{rgb}{0.55,0.55,0.55}
\definecolor{grey56}{rgb}{0.56,0.56,0.56}
\definecolor{grey57}{rgb}{0.57,0.57,0.57}
\definecolor{grey58}{rgb}{0.58,0.58,0.58}
\definecolor{grey59}{rgb}{0.59,0.59,0.59}
\definecolor{grey5}{rgb}{0.05,0.05,0.05}
\definecolor{grey60}{rgb}{0.60,0.60,0.60}
\definecolor{grey61}{rgb}{0.61,0.61,0.61}
\definecolor{grey62}{rgb}{0.62,0.62,0.62}
\definecolor{grey63}{rgb}{0.63,0.63,0.63}
\definecolor{grey64}{rgb}{0.64,0.64,0.64}
\definecolor{grey65}{rgb}{0.65,0.65,0.65}
\definecolor{grey66}{rgb}{0.66,0.66,0.66}
\definecolor{grey67}{rgb}{0.67,0.67,0.67}
\definecolor{grey68}{rgb}{0.68,0.68,0.68}
\definecolor{grey69}{rgb}{0.69,0.69,0.69}
\definecolor{grey6}{rgb}{0.06,0.06,0.06}
\definecolor{grey70}{rgb}{0.70,0.70,0.70}
\definecolor{grey71}{rgb}{0.71,0.71,0.71}
\definecolor{grey72}{rgb}{0.72,0.72,0.72}
\definecolor{grey73}{rgb}{0.73,0.73,0.73}
\definecolor{grey74}{rgb}{0.74,0.74,0.74}
\definecolor{grey75}{rgb}{0.75,0.75,0.75}
\definecolor{grey76}{rgb}{0.76,0.76,0.76}
\definecolor{grey77}{rgb}{0.77,0.77,0.77}
\definecolor{grey78}{rgb}{0.78,0.78,0.78}
\definecolor{grey79}{rgb}{0.79,0.79,0.79}
\definecolor{grey7}{rgb}{0.07,0.07,0.07}
\definecolor{grey80}{rgb}{0.80,0.80,0.80}
\definecolor{grey81}{rgb}{0.81,0.81,0.81}
\definecolor{grey82}{rgb}{0.82,0.82,0.82}
\definecolor{grey83}{rgb}{0.83,0.83,0.83}
\definecolor{grey84}{rgb}{0.84,0.84,0.84}
\definecolor{grey85}{rgb}{0.85,0.85,0.85}
\definecolor{grey86}{rgb}{0.86,0.86,0.86}
\definecolor{grey87}{rgb}{0.87,0.87,0.87}
\definecolor{grey88}{rgb}{0.88,0.88,0.88}
\definecolor{grey89}{rgb}{0.89,0.89,0.89}
\definecolor{grey8}{rgb}{0.08,0.08,0.08}
\definecolor{grey90}{rgb}{0.90,0.90,0.90}
\definecolor{grey91}{rgb}{0.91,0.91,0.91}
\definecolor{grey92}{rgb}{0.92,0.92,0.92}
\definecolor{grey93}{rgb}{0.93,0.93,0.93}
\definecolor{grey94}{rgb}{0.94,0.94,0.94}
\definecolor{grey95}{rgb}{0.95,0.95,0.95}
\definecolor{grey96}{rgb}{0.96,0.96,0.96}
\definecolor{grey97}{rgb}{0.97,0.97,0.97}
\definecolor{grey98}{rgb}{0.98,0.98,0.98}
\definecolor{grey99}{rgb}{0.99,0.99,0.99}
\definecolor{grey9}{rgb}{0.09,0.09,0.09}
\definecolor{grey}{rgb}{0.75,0.75,0.75}
\definecolor{honeydew1}{rgb}{0.94,1.00,0.94}
\definecolor{honeydew2}{rgb}{0.88,0.93,0.88}
\definecolor{honeydew3}{rgb}{0.76,0.80,0.76}
\definecolor{honeydew4}{rgb}{0.51,0.55,0.51}
\definecolor{honeydew}{rgb}{0.94,1.00,0.94}
\definecolor{hotpink}{rgb}{1.00,0.41,0.71}
\definecolor{indianred}{rgb}{0.80,0.36,0.36}
\definecolor{ivory1}{rgb}{1.00,1.00,0.94}
\definecolor{ivory2}{rgb}{0.93,0.93,0.88}
\definecolor{ivory3}{rgb}{0.80,0.80,0.76}
\definecolor{ivory4}{rgb}{0.55,0.55,0.51}
\definecolor{ivory}{rgb}{1.00,1.00,0.94}
\definecolor{khaki1}{rgb}{1.00,0.96,0.56}
\definecolor{khaki2}{rgb}{0.93,0.90,0.52}
\definecolor{khaki3}{rgb}{0.80,0.78,0.45}
\definecolor{khaki4}{rgb}{0.55,0.53,0.31}
\definecolor{khaki}{rgb}{0.94,0.90,0.55}
\definecolor{lavenderblush}{rgb}{1.00,0.94,0.96}
\definecolor{lavender}{rgb}{0.90,0.90,0.98}
\definecolor{lawngreen}{rgb}{0.49,0.99,0.00}
\definecolor{lemonchiffon}{rgb}{1.00,0.98,0.80}
\definecolor{lightblue}{rgb}{0.68,0.85,0.90}
\definecolor{lightcoral}{rgb}{0.94,0.50,0.50}
\definecolor{lightcyan}{rgb}{0.88,1.00,1.00}
\definecolor{lightgoldenrod}{rgb}{0.93,0.87,0.51}
\definecolor{lightgoldenrod}{rgb}{0.98,0.98,0.82}
\definecolor{lightgray}{rgb}{0.83,0.83,0.83}
\definecolor{lightgreen}{rgb}{0.56,0.93,0.56}
\definecolor{lightgrey}{rgb}{0.83,0.83,0.83}
\definecolor{lightpink}{rgb}{1.00,0.71,0.76}
\definecolor{lightsalmon}{rgb}{1.00,0.63,0.48}
\definecolor{lightsea}{rgb}{0.13,0.70,0.67}
\definecolor{lightsky}{rgb}{0.53,0.81,0.98}
\definecolor{lightslate}{rgb}{0.47,0.53,0.60}
\definecolor{lightslate}{rgb}{0.47,0.53,0.60}
\definecolor{lightslate}{rgb}{0.52,0.44,1.00}
\definecolor{lightsteel}{rgb}{0.69,0.77,0.87}
\definecolor{lightyellow}{rgb}{1.00,1.00,0.88}
\definecolor{limegreen}{rgb}{0.20,0.80,0.20}
\definecolor{linen}{rgb}{0.98,0.94,0.90}
\definecolor{magenta1}{rgb}{1.00,0.00,1.00}
\definecolor{magenta2}{rgb}{0.93,0.00,0.93}
\definecolor{magenta3}{rgb}{0.80,0.00,0.80}
\definecolor{magenta4}{rgb}{0.55,0.00,0.55}
\definecolor{magenta}{rgb}{1.00,0.00,1.00}
\definecolor{maroon1}{rgb}{1.00,0.20,0.70}
\definecolor{maroon2}{rgb}{0.93,0.19,0.65}
\definecolor{maroon3}{rgb}{0.80,0.16,0.56}
\definecolor{maroon4}{rgb}{0.55,0.11,0.38}
\definecolor{maroon}{rgb}{0.69,0.19,0.38}
\definecolor{mediumaquamarine}{rgb}{0.40,0.80,0.67}
\definecolor{mediumblue}{rgb}{0.00,0.00,0.80}
\definecolor{mediumorchid}{rgb}{0.73,0.33,0.83}
\definecolor{mediumpurple}{rgb}{0.58,0.44,0.86}
\definecolor{mediumsea}{rgb}{0.24,0.70,0.44}
\definecolor{mediumslate}{rgb}{0.48,0.41,0.93}
\definecolor{mediumspring}{rgb}{0.00,0.98,0.60}
\definecolor{mediumturquoise}{rgb}{0.28,0.82,0.80}
\definecolor{mediumviolet}{rgb}{0.78,0.08,0.52}
\definecolor{midnightblue}{rgb}{0.10,0.10,0.44}
\definecolor{mintcream}{rgb}{0.96,1.00,0.98}
\definecolor{mistyrose}{rgb}{1.00,0.89,0.88}
\definecolor{moccasin}{rgb}{1.00,0.89,0.71}
\definecolor{navajowhite}{rgb}{1.00,0.87,0.68}
\definecolor{navyblue}{rgb}{0.00,0.00,0.50}
\definecolor{navy}{rgb}{0.00,0.00,0.50}
\definecolor{oldlace}{rgb}{0.99,0.96,0.90}
\definecolor{olivedrab}{rgb}{0.42,0.56,0.14}
\definecolor{orange1}{rgb}{1.00,0.65,0.00}
\definecolor{orange2}{rgb}{0.93,0.60,0.00}
\definecolor{orange3}{rgb}{0.80,0.52,0.00}
\definecolor{orange4}{rgb}{0.55,0.35,0.00}
\definecolor{orangered}{rgb}{1.00,0.27,0.00}
\definecolor{orange}{rgb}{1.00,0.65,0.00}
\definecolor{orchid1}{rgb}{1.00,0.51,0.98}
\definecolor{orchid2}{rgb}{0.93,0.48,0.91}
\definecolor{orchid3}{rgb}{0.80,0.41,0.79}
\definecolor{orchid4}{rgb}{0.55,0.28,0.54}
\definecolor{orchid}{rgb}{0.85,0.44,0.84}
\definecolor{palegoldenrod}{rgb}{0.93,0.91,0.67}
\definecolor{palegreen}{rgb}{0.60,0.98,0.60}
\definecolor{paleturquoise}{rgb}{0.69,0.93,0.93}
\definecolor{paleviolet}{rgb}{0.86,0.44,0.58}
\definecolor{papayawhip}{rgb}{1.00,0.94,0.84}
\definecolor{peachpuff}{rgb}{1.00,0.85,0.73}
\definecolor{peru}{rgb}{0.80,0.52,0.25}
\definecolor{pink1}{rgb}{1.00,0.71,0.77}
\definecolor{pink2}{rgb}{0.93,0.66,0.72}
\definecolor{pink3}{rgb}{0.80,0.57,0.62}
\definecolor{pink4}{rgb}{0.55,0.39,0.42}
\definecolor{pink}{rgb}{1.00,0.75,0.80}
\definecolor{plum1}{rgb}{1.00,0.73,1.00}
\definecolor{plum2}{rgb}{0.93,0.68,0.93}
\definecolor{plum3}{rgb}{0.80,0.59,0.80}
\definecolor{plum4}{rgb}{0.55,0.40,0.55}
\definecolor{plum}{rgb}{0.87,0.63,0.87}
\definecolor{powderblue}{rgb}{0.69,0.88,0.90}
\definecolor{purple1}{rgb}{0.61,0.19,1.00}
\definecolor{purple2}{rgb}{0.57,0.17,0.93}
\definecolor{purple3}{rgb}{0.49,0.15,0.80}
\definecolor{purple4}{rgb}{0.33,0.10,0.55}
\definecolor{purple}{rgb}{0.63,0.13,0.94}
\definecolor{red1}{rgb}{1.00,0.00,0.00}
\definecolor{red2}{rgb}{0.93,0.00,0.00}
\definecolor{red3}{rgb}{0.80,0.00,0.00}
\definecolor{red4}{rgb}{0.55,0.00,0.00}
\definecolor{red}{rgb}{1.00,0.00,0.00}
\definecolor{rosybrown}{rgb}{0.74,0.56,0.56}
\definecolor{royalblue}{rgb}{0.25,0.41,0.88}
\definecolor{saddlebrown}{rgb}{0.55,0.27,0.07}
\definecolor{salmon1}{rgb}{1.00,0.55,0.41}
\definecolor{salmon2}{rgb}{0.93,0.51,0.38}
\definecolor{salmon3}{rgb}{0.80,0.44,0.33}
\definecolor{salmon4}{rgb}{0.55,0.30,0.22}
\definecolor{salmon}{rgb}{0.98,0.50,0.45}
\definecolor{sandybrown}{rgb}{0.96,0.64,0.38}
\definecolor{seagreen}{rgb}{0.18,0.55,0.34}
\definecolor{seashell1}{rgb}{1.00,0.96,0.93}
\definecolor{seashell2}{rgb}{0.93,0.90,0.87}
\definecolor{seashell3}{rgb}{0.80,0.77,0.75}
\definecolor{seashell4}{rgb}{0.55,0.53,0.51}
\definecolor{seashell}{rgb}{1.00,0.96,0.93}
\definecolor{sienna1}{rgb}{1.00,0.51,0.28}
\definecolor{sienna2}{rgb}{0.93,0.47,0.26}
\definecolor{sienna3}{rgb}{0.80,0.41,0.22}
\definecolor{sienna4}{rgb}{0.55,0.28,0.15}
\definecolor{sienna}{rgb}{0.63,0.32,0.18}
\definecolor{skyblue}{rgb}{0.53,0.81,0.92}
\definecolor{slateblue}{rgb}{0.42,0.35,0.80}
\definecolor{slategray}{rgb}{0.44,0.50,0.56}
\definecolor{slategrey}{rgb}{0.44,0.50,0.56}
\definecolor{snow1}{rgb}{1.00,0.98,0.98}
\definecolor{snow2}{rgb}{0.93,0.91,0.91}
\definecolor{snow3}{rgb}{0.80,0.79,0.79}
\definecolor{snow4}{rgb}{0.55,0.54,0.54}
\definecolor{snow}{rgb}{1.00,0.98,0.98}
\definecolor{springgreen}{rgb}{0.00,1.00,0.50}
\definecolor{steelblue}{rgb}{0.27,0.51,0.71}
\definecolor{tan1}{rgb}{1.00,0.65,0.31}
\definecolor{tan2}{rgb}{0.93,0.60,0.29}
\definecolor{tan3}{rgb}{0.80,0.52,0.25}
\definecolor{tan4}{rgb}{0.55,0.35,0.17}
\definecolor{tan}{rgb}{0.82,0.71,0.55}
\definecolor{thistle1}{rgb}{1.00,0.88,1.00}
\definecolor{thistle2}{rgb}{0.93,0.82,0.93}
\definecolor{thistle3}{rgb}{0.80,0.71,0.80}
\definecolor{thistle4}{rgb}{0.55,0.48,0.55}
\definecolor{thistle}{rgb}{0.85,0.75,0.85}
\definecolor{tomato1}{rgb}{1.00,0.39,0.28}
\definecolor{tomato2}{rgb}{0.93,0.36,0.26}
\definecolor{tomato3}{rgb}{0.80,0.31,0.22}
\definecolor{tomato4}{rgb}{0.55,0.21,0.15}
\definecolor{tomato}{rgb}{1.00,0.39,0.28}
\definecolor{turquoise1}{rgb}{0.00,0.96,1.00}
\definecolor{turquoise2}{rgb}{0.00,0.90,0.93}
\definecolor{turquoise3}{rgb}{0.00,0.77,0.80}
\definecolor{turquoise4}{rgb}{0.00,0.53,0.55}
\definecolor{turquoise}{rgb}{0.25,0.88,0.82}
\definecolor{violetred}{rgb}{0.82,0.13,0.56}
\definecolor{violet}{rgb}{0.93,0.51,0.93}
\definecolor{wheat1}{rgb}{1.00,0.91,0.73}
\definecolor{wheat2}{rgb}{0.93,0.85,0.68}
\definecolor{wheat3}{rgb}{0.80,0.73,0.59}
\definecolor{wheat4}{rgb}{0.55,0.49,0.40}
\definecolor{wheat}{rgb}{0.96,0.87,0.70}
\definecolor{whitesmoke}{rgb}{0.96,0.96,0.96}
\definecolor{white}{rgb}{1.00,1.00,1.00}
\definecolor{yellow1}{rgb}{1.00,1.00,0.00}
\definecolor{yellow2}{rgb}{0.93,0.93,0.00}
\definecolor{yellow3}{rgb}{0.80,0.80,0.00}
\definecolor{yellow4}{rgb}{0.55,0.55,0.00}
\definecolor{yellowgreen}{rgb}{0.60,0.80,0.20}
\definecolor{yellow}{rgb}{1.00,1.00,0.00}
\usepackage{amsfonts}
\usepackage{longtable}
\usepackage{bm}
\usepackage{longtable}
\usepackage{multirow}
\usepackage{pifont}
\usepackage{gensymb}
\newcommand{\iris}{{\em IRIS}}
\def \mgii  {Mg\,{\sc ii}}
\def \caii  {Ca\,{\sc ii}}
\def \cii  {C\,{\sc ii}}
\def \fexxi {Fe\,{\sc xxi}}
\def \oiv  {O\,{\sc iv}}
\def \hei  {He\,{\sc i}}
\def \siiv  {Si\,{\sc iv}}
\def \si  {S\,{\sc i}}
\usepackage{booktabs} 

\newcommand{\gskfont}{
  \bfseries
  \color{applered}
}
\definecolor{applered}{rgb}{0.89, 0.02, 0.17}
\DeclareTextFontCommand{\gsk}{\gskfont}

\newcommand{\radyn}{\texttt{RADYN}}
\newcommand{\fpcode}{\texttt{FP}}
\newcommand{\radynfp}{\texttt{RADYN+FP}}
\newcommand{\rhtiago}{\texttt{RH15D}}

\begin{document}
\title{Solar Flare Ribbon Fronts I: Constraining flare energy deposition with IRIS spectroscopy}
  \author{Vanessa Polito}
  	\email{polito@baeri.org}
	\affiliation{Bay Area Environmental Research Institute, NASA Research Park,  Moffett Field, CA 94035-0001, USA}
        \affiliation{Lockheed Martin Solar and Astrophysics Laboratory, Building 252, 3251 Hanover Street, Palo Alto, CA 94304, USA}
	\affiliation{Department of Physics, Oregon State University, 301 Weniger Hall, Corvallis, OR 97331}
\author{Graham~S. Kerr}

	\affil{NASA Goddard Space Flight Center, Heliophysics Sciences Division, Code 671, 8800 Greenbelt Rd., Greenbelt, MD 20771, USA}
 	\affil{Department of Physics, Catholic University of America, 620 Michigan Avenue, Northeast, Washington, DC 20064, USA}
 
	 \author{Yan Xu}
	 \affil{Institute for Space Weather Sciences, New Jersey Institute of Technology, 323 Martin Luther King Boulevard, Newark, NJ 07102-1982}
	 \affil{Big Bear Solar Observatory, New Jersey Institute of Technology, 40386 North Shore Lane, Big Bear City, CA 92314-9672, USA}

	\author{Viacheslav~M. Sadykov}
	\affil{Physics \& Astronomy Department, Georgia State University, 25 Park Place NE, Atlanta, GA 30303, USA}
	
 \author{Juraj Lorincik}
	\affiliation{Bay Area Environmental Research Institute, NASA Research Park,  Moffett Field, CA 94035-0001, USA}
        \affiliation{Lockheed Martin Solar and Astrophysics Laboratory, Building 252, 3251 Hanover Street, Palo Alto, CA 94304, USA}

\begin{abstract}
   
Lower atmospheric lines show peculiar profiles at the leading edge of ribbons during solar flares. In particular, increased absorption of the BBSO/GST \hei~10830~\AA\ line  \citep[e.g.][]{Xu2016}, as well as broad and centrally reversed profiles in the spectra of the \mgii~and \cii~lines observed by the \iris~satellite \citep[e.g.][]{Panos2018,Panos2021a} have been reported. In this work, we aim to understand the  physical origin of the \iris\ ribbon front line profiles, which seem to be common of many, if not all, flares. To achieve this, we quantify the spectral properties of the \iris~\mgii~ribbon front profiles during four large flares and perform a detailed comparison with a grid of radiative hydrodynamic models using the \radynfp~code. We also studied their transition region counterparts, finding that these ribbon front locations are regions where transition region emission and chromospheric evaporation are considerably weaker compared to other parts of the ribbons. Based on our comparison between the \iris~observations and modelling, our interpretation is that there are different heating regimes at play in the leading and trailing regions of the ribbons. More specifically, we suggest that bombardment of the chromosphere by more gradual and modest non-thermal electron  energy fluxes can qualitatively explain the \iris~observations at the ribbon front, while stronger and more impulsive energy fluxes are required to drive chromospheric evaporation and more intense TR emission. Our results provide a possible physical origin for the peculiar behaviour of the \iris~chromospheric lines in the ribbon leading edge and new  constraints for the flare models.

  \end{abstract}

\section{Introduction}
\label{Sect:intro}

During solar flares magnetic reconnection liberates energy from the stressed coronal magnetic field \citep{2002A&ARv..10..313P,2013A&A...555A..77J}. This energy manifests in several forms, including the acceleration of large amounts of particles, but is ultimately radiated away \citep[e.g.][]{2012ApJ...759...71E}. Flares are characterised by this intense broadband enhancement to the solar radiative output, and it is through careful study of that radiation that we can extract information about the magnetic reconnection, energy release, and particle acceleration processes that occur during flares. 

Flare energy is carried by some agent from the release site in corona to the lower atmosphere (the chromosphere and transition region, TR), where it produces ribbon-like structures observable in the UV, optical and near-infrared \citep[e.g.][]{2011SSRv..159...19F}. This agent is typically thought to be non-thermal electrons, due to the almost ubiquitous presence of compact hard X-ray (HXR) sources that are spatially associated with chromospheric/TR ribbons \citep{2011SSRv..159..301K,2011SSRv..159..107H,2011SSRv..159...19F}. Other mechanisms that are likely acting include non-thermal protons or heavier ions \cite[e.g.][]{2000AIPC..522..401R,2012ApJ...759...71E}, the conductive heat flux resulting from direct heating in the corona \citep[e.g.][]{2012ApJ...754...54B,2021ApJ...912...25A}, or magnetohydrodynamic (MHD) waves such as Alfv\'en waves \citep[e.g.][]{2008ApJ...675.1645F,2016ApJ...827..101K,2016ApJ...818L..20R,2018ApJ...853..101R}, though these alternative mechanisms are not as well characterised as the `electron beam' model. See also the discussion in Section 5.4 of \cite{2022ApJ...926...53C}.  

While HXR observations to-date offer relatively coarse spatial resolution \citep[e.g. the Reuven Ramaty High Energy Solar Spectroscopic Imager, RHESSI, had a spatial resolution of $\sim 2.3^{\prime\prime}$ up to 100~keV, and $\sim7^{\prime\prime}$ up to 400~keV,][]{2002SoPh..210....3L}, there is now a wealth of high spatial resolution observations of the lower atmosphere in the optical, UV and near-infrared, both ground and space-based. Two examples relevant for the research discussed in this manuscript are the Interface Region Imaging Spectrograph \citep[IRIS;][]{DePontieu2014}, and the Goode Solar Telescope at the Big Bear Solar Telescope \citep[BBSO/GST;][]{2012ASPC..463..357G}. IRIS offers spatial resolution of $0.3-0.4^{\prime\prime}$ in the far- and near-UV (FUV \& NUV), providing both images and spectra. BBSO/GST also provides imaging and spectra, but in the optical and near-infrared (NIR), where the diffraction limit of its $1.6$~m telescope is $0.08^{\prime\prime}$ at $500$~nm, and $0.16^{\prime\prime}$ at $1$~micron.  These high-resolution observations have revealed in recent years that the narrow leading edge of flare ribbons exhibits features that differ from the brighter trailing portions of the ribbons. 

Ribbon leading edges (or ribbon fronts) are the footpoints of the newest reconnected flare loops and therefore represent the site of initial energy deposition, carry vital information about the energy transport and dissipation mechanisms, and thus the energy release and particle acceleration processes themselves. 

We focus on two examples of ribbon front behaviour that offer scope for fruitful diagnostics of energy deposition: the dimming of ribbons observed in \hei~ 10830\AA, and the presence of unique spectral shapes of the \mgii~near-UV (NUV) spectral lines. 

Spectral lines of orthohelium (the \hei~10830\AA\ and \hei~ D3 lines) have been observed to, curiously, undergo periods of \textsl{dimming} during solar flares, before brightening \citep[e.g.][]{1980ApJ...235..618Z,2013ApJ...774...60L,Xu2016,2022ApJ...924L..18X,2018JASTP.173...50K}. This has been seen both in imaging and spectroscopy \citep{Xu2016,2022ApJ...924L..18X}. Recent BBSO/GST observations showed that some of these dimmings occur along the leading edge of propagating flare ribbons \citep{Xu2016,2022ApJ...924L..18X}, i.e. immediately following the injection of flare energy. These `negative' flare ribbons persisted for several dozens of seconds to over a minute, with a width around $\sim350-500$~km.  Also, in the two-ribbon flares studied by \cite{Xu2016}, only one ribbon in each flare exhibited the dimming, from which we can infer that energy deposition into each ribbon differed in some regard.

Clearly such observations suggest that the initial energy deposition into the chromosphere produced a response that differs from the typical expectation (that is, rapid impulsive brightening). Two suggestions were made: (1) that enhanced extreme-UV (EUV) radiation from the flare heated corona increased the photoionisation rate of \hei, with subsequent recombinations to orthohelium and increased opacity that absorbed photospheric radiation (the photoionisation-recombination mechanism, PRM), or (2) that non-thermal electrons within the beam collisionally ionised \hei, that subsequently recombined and overpopulated orthohelium sufficiently to absorb more photospheric radiation (the collisional ionisation-recomibation mechanism, CRM). 

Using field-aligned radiation hydrodynamic (RHD) modelling, \cite{2020ApJ...897L...6H} and \cite{2021ApJ...912..153K} showed that electron beam driven flare simulations could produce the observed pattern of dimming followed by brightening. \cite{2021ApJ...912..153K} demonstrated that simulations that only included the PRM were unable to produce dimming of \hei~but simulations that also included non-thermal collisional ionisation of \hei~ were successful in producing dimming of the 10830\AA\ line at flare onset. The characteristics of that dimming were related to the properties of the injected non-thermal electron distribution. 

The results of \cite{2021ApJ...912..153K} suggest that where we observe dimming of \hei~10830 ribbon fronts, followed by brightenings, non-thermal particles are present in the chromosphere. A harder non-thermal electron energy spectrum (larger proportion of higher energy non-thermal electrons compared to lower energy non-thermal electrons) and weaker flux of those electrons resulted in stronger, more sustained dimming. However, the lifetime of the dimming (i.e. the time during which a particular area existed as a `ribbon front' source) was not consistent with observations. We were only able to model enhanced absorption for a few seconds, compared to several dozen seconds observed by \cite{Xu2016}.

Routine high resolutions observations of the Sun in the FUV and NUV have been available since the launch of IRIS in 2013, and since then many hundreds of flares have been observed. One of the strongest sets of lines observed by IRIS are the \mgii~h \& k resonance lines, and the \mgii~subordinate triplet. Forming over a range of chromospheric altitudes, all together these lines are diagnostically important \citep[see for example the quiet Sun diagnostics of][]{2013ApJ...772...89L,2013ApJ...772...90L,Pereira2013}. Modern inversion codes coupled with machine learning techniques also mean that it is possible to estimate the atmospheric stratification of temperature, electron density and other plasma properties from these lines \citep[][]{2019ApJ...875L..18S}.

In flares, however, the \mgii~ lines appear very different from their quiescent counterparts \citep[e.g.][]{2015A&A...582A..50K,2015SoPh..290.3525L,Panos2018,2015ApJ...807L..22G,2020ApJ...895....6G}, making extracting the information they carry more troublesome. Sources within flare ribbons are typically single-peaked (in contrast to the central reversal commonplace elsewhere), broadened, very intense, redshifted or with marked wing asymmetries, and show non-Gaussian line wings (they can appear quite Lorentzian). At the same time, the subordinate lines go into emission, and the line ratios can be observed to decrease slightly (likely due to some opacity changes).

Attempts to model these lines in flares \citep[e.g.][]{2016ApJ...827..101K, 2019ApJ...883...57K, 2019ApJ...885..119K, 2019ApJ...879...19Z, 2017ApJ...842...82R, 2019ApJ...878L..15H} have shown that high densities are required to `fill in' the central reversal, that chromospheric condensations can produce redshifts and asymmetric flows, and that turbulent broadening is likely not the main culprit behind the excess line widths (though we are still unable to sufficiently broaden the lines comparable to observations). During flares, the subordinate lines have also been found to form higher in altitude, forming close to the resonance lines \citep[][]{2019ApJ...871...23K, 2019ApJ...879...19Z} so that their being in emission is not necessarily a sign of deep heating as is the case in the quiet Sun \citep[e.g.][]{2015ApJ...806...14P}.

Since \iris~has observed hundreds of thousands of individual \mgii~ spectra from many flares, it is advantageous to perform clustering techniques such as $k$-means in order to sift through this vast dataset and identify commonalities that might otherwise be missed. This was performed initially for one flare by \cite{Xu2016} who noted that ribbon front \mgii~ profiles showed marked differences compared to the brighter portions of the ribbon. This was greatly expanded upon by \cite{Panos2018} who analysed 33 M and X class flares. They found that in addition to the single peaked `flare' profiles described above, there was a class of \mgii\ profiles that were located at the leading edge of some propagating ribbon sources \cite[with variable lifetimes, but on the order of 1-3 minutes][]{Panos2021a, Panos2021b}. Those ribbon front profiles had deep central reversals, slightly blueshifted cores, were extremely broad, and showed subordinate lines in emission. While \cite{Panos2018} and \cite{2018PASJ...70..100T} (who observed similar features in a C class flare kernel) speculate that these could be caused by superposition of very strong unresolved flows at different chromospheric temperatures, enhanced turbulence at the leading edge of flare ribbons, or due to rising cool chromospheric material, there has been no self-consistent flare modelling that has explained these observations.  

Finally we briefly note other behaviours that illustrate the importance of studying flare ribbon fronts, and ribbons generally, in order to understand fundamental flare processes. Using \iris~slit-jaw imager (SJI) images in the FUV, \cite{2022ApJ...926..218N} found that while the ribbons are globally laminar, they contain fine scale structure in both space and time. Ribbon front widths varied over time, and were seen to activate some 1-3 minutes (in an average sense) before reaching peak intensity. This fine scale structuring may be related directly to dynamics in the current sheet itself. Similarly, \cite{2021ApJ...922..117F} used very high-cadence (1.7~s) IRIS data to relate flare ribbon dynamics to current sheet instabilities. From that same dataset, \cite{Jeffreyeaav2794} previously demonstrated that \siiv~ line widths increased prior to a strong increase in line intensity. The build up of MHD turbulence during the early phases of ribbon development was posited as an explanation.\\

Here we continue our exploration of flare ribbon fronts that we started in \cite{2021ApJ...912..153K}, by determining if the same models that could produce \hei\ 10830\AA\ dimming can produce \mgii~NUV spectra consistent with IRIS ribbon front observations. To facilitate that comparison we put the characteristics of the \mgii~NUV ribbon front spectra on a more quantitative footing, building metrics that describe the centroid shifts, central reversal depths, peak asymmetry, amongst others. 

In Section~\ref{Sect:obs} we quantify the characteristics of \mgii~ ribbon fronts observed by IRIS, and in Section~\ref{Sect:sims} we synthesise \mgii~NUV spectra, before performing a model-data comparison of the line metrics in Section~\ref{Sect:discussion}. We do not address the line formation properties or ribbon front lifetimes here, leaving that to a follow on work (Kerr et al, \textsl{In Prep.}).

\begin{figure*}[!ht]
\center
\includegraphics[width=\textwidth]{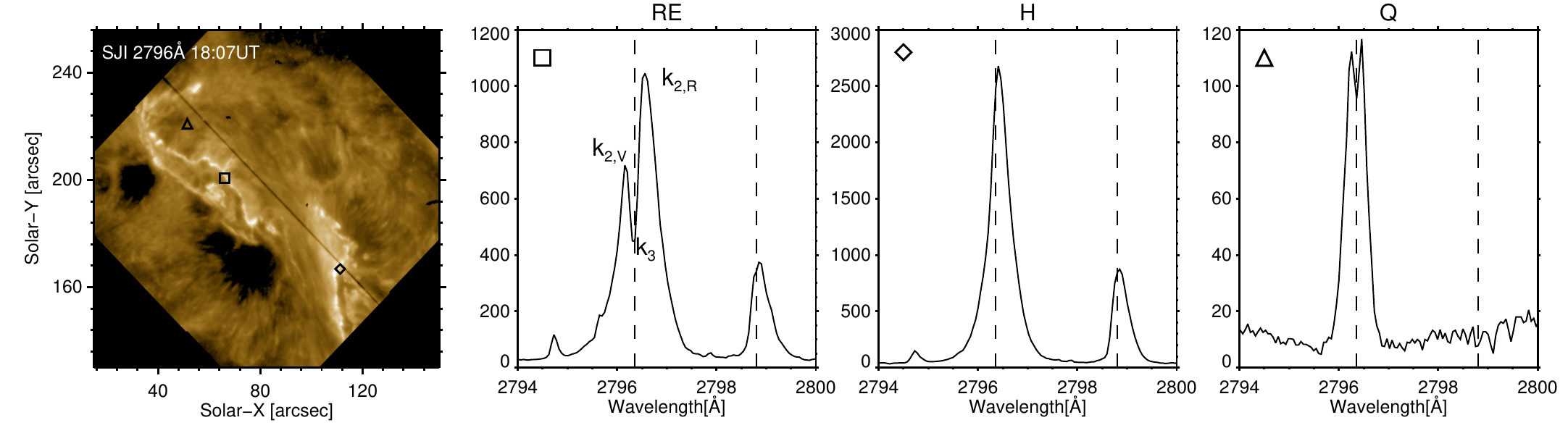}
\caption{Overview of different types of \mgii~profiles and their location within the flare (panel a). Example of "ribbon-edge" (RE, panel b) profiles which are typically found in the leading edge of flare ribbons \citep[e.g.][]{Panos2018}; "heating" single peaked profiles (H, panel c), which are more typical of the trailing edge of the ribbon, and a "quiet" non-flaring (Q, panel d) profile for comparison. On the RE panel we note the common definition of the \mgii\ profiles k3 (line core) and k2v \& k2r (flanking emission peaks on the blue and red side of the line core).}
\label{Fig:profiles}
\end{figure*}

\section{IRIS observations of flare ribbons}
\label{Sect:obs}
In this section we describe the methodology we use to quantify the spectral characteristics of the \iris~\mgii~ribbon leading edge (RE, Figure~\ref{Fig:profiles}) profiles, that we will compare with the predictions from the radiation-hydrodynamic models in Sect.~\ref{Sect:sims}. Since its launch in 2013, \iris~has provided an unprecedented view of the solar atmosphere from the photosphere to the flaring corona \citep{DePontieu2014,DePontieu2021}. The satellite consists of a spectrograph and a Slit-Jaw Imager (SJI) observing the Sun in both the far  ultraviolet (FUV) and near ultraviolet (NUV) ranges at unprecedented spatial (0.33--0.4\arcsec), temporal (down to 1s or less) and spectral (2.7 km s$^{-1}$ pixels) resolution. The \iris~spectrograph observes continua and spectral lines formed over a broad range of temperatures (logT[K]~$\approx$~3.5--7) in both sit-and-stare and raster modes. The \iris~rasters can be: (1) dense, if the raster step size is the same as the slit width (0.33\arcsec), (2) sparse if the step size is 1\arcsec, or (3) coarse if the step size is 2\arcsec.
\begin{figure*}
\center
\includegraphics[width=\textwidth]{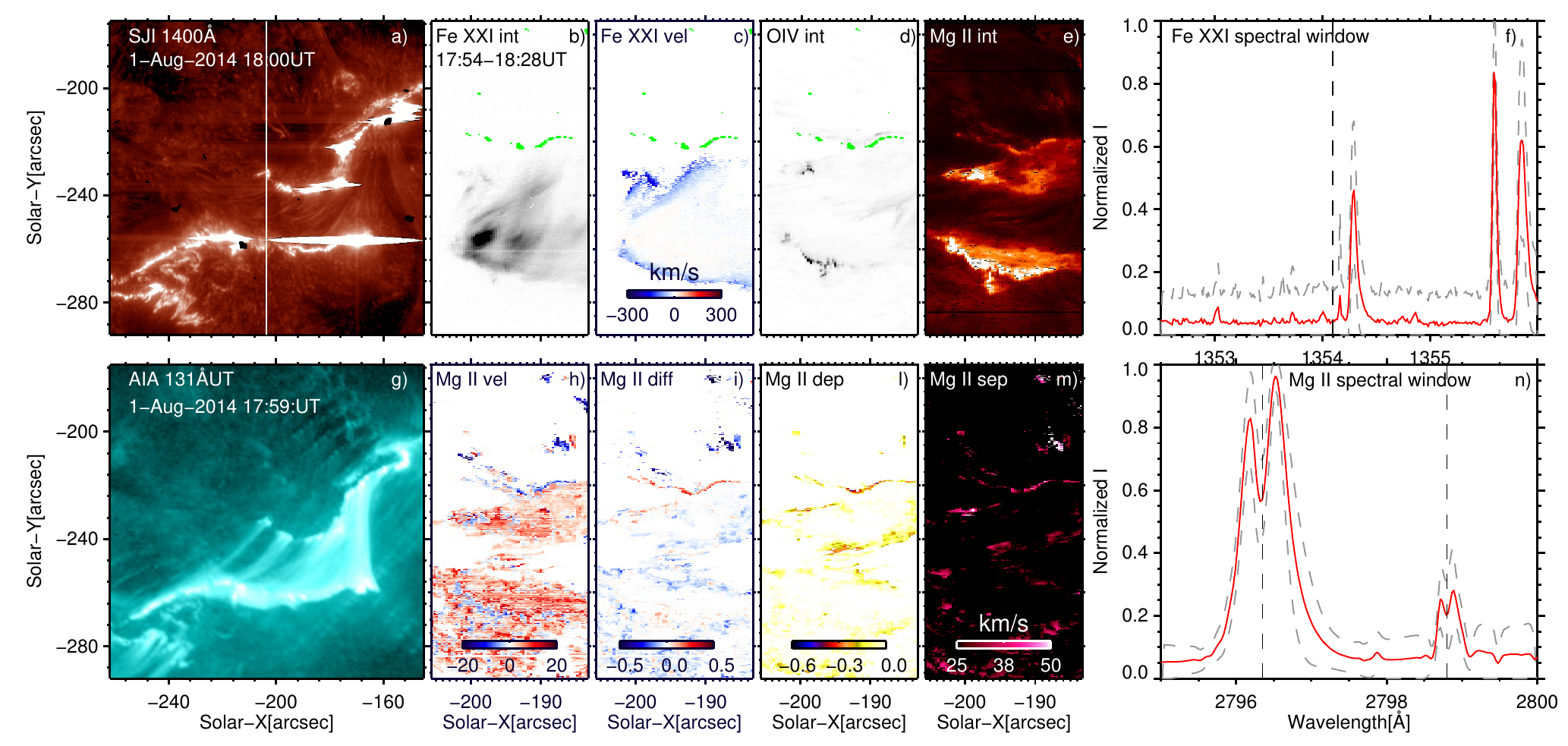}
\caption{Overview of FL1: \iris~SJI 1400\AA~image with raster FOV overlaid (a); \fexxi~intensity (b) and velocity (c), \oiv~intensity (d) from single Gaussian fits with overlaid the contours of the pixels with the RE profiles in green color; \mgii~k3 intensity (e) and velocity (h), k2 peak difference (i), depth of central reversal (l) and k2 peak separation (m), see formulae in the main text. Panels f) and n) show the average spectra (red line) $\pm$ the standard deviation (grey dotted line) in the RE pixel locations for the \fexxi~and \mgii~k windows respectively. Panel g) shows an overview of the flare in the AIA~131\AA~filter. The time in panel b) refers to the start and end time of the raster.} 
\label{Fig:flare_1}
\end{figure*} 

\begin{figure*}
\center
\includegraphics[width=\textwidth]{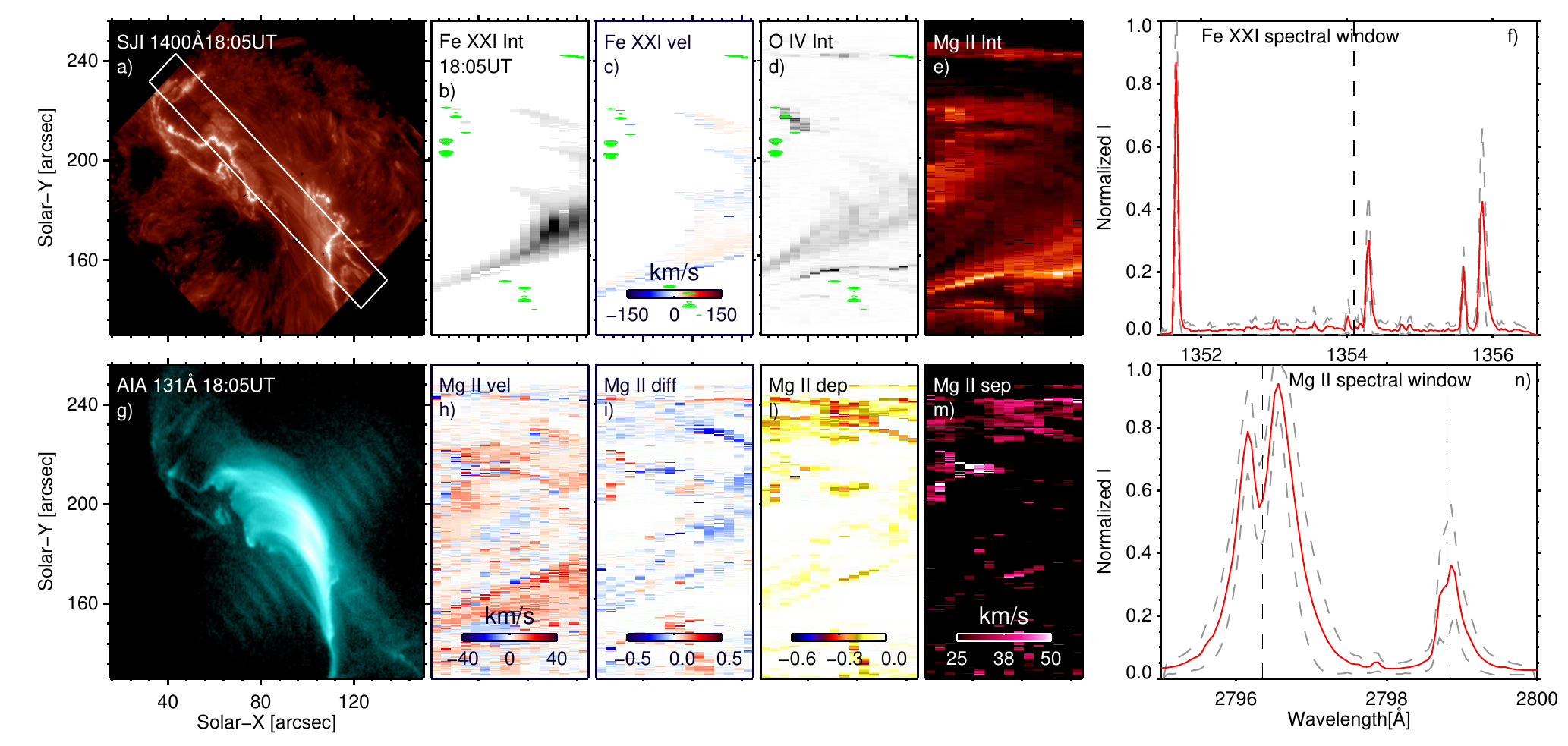}
\caption{Overview of FL2 (one example raster): for the panels' description, see Figure~\ref{Fig:flare_1} and Sect.~\ref{Sect:obs}. This observation had a rotation angle of 45 degrees, but the raster data in panels b)--e) and h)--m) are not rotated for convenience. The time in panel b) refers to the mid-time of the example raster (the raster cadence is $\approx$~32s). A movie associated with this figure is also available. } 
\label{Fig:flare_2}
\end{figure*}

We analyse a sample of four different flares observed by \iris: 
\begin{itemize}
    \item \textbf{Flare 1}: 2014-08-01 M-class flare (FL1, Figure~\ref{Fig:flare_1}): Large and dense 64-step raster with cadence of $\approx$~33~minutes, exposure time of $\approx$~30~s and step cadence of $\approx$~32~s. This flare was analysed by \citet{Xu2016}, who observed enhanced absorption in the GST \hei~line and strongly reversed \iris~\mgii~profiles at the edge of one of the ribbons (see also Section~\ref{Sect:intro}). We analyse the same raster as that in \citet{Xu2016}, from $\approx$~15:55 to 18:28UT, where the RE profiles are observed. 
    \item \textbf{Flare 2}: 2015-06-22 M-class flare (FL2, Figure~\ref{Fig:flare_2}): Large and sparse 16-step raster with raster cadence of $\approx$~30~s, average exposure time of $\approx$~1~s (down to $\approx$~0.4~s for the NUV lines) with step cadence of $\approx$~2~s. The spectrograph's observation has a spatial and spectral (for both the FUV and NUV channels) binning of 2 and is rotated by 45 degrees. We analyse several rasters covering about 20 minutes around the flare's peak time, from $\approx$~17:51 to 18:10UT. 
    \item \textbf{Flare 3}: 2014-09-10 X-class flare (FL3, Figure~\ref{Fig:flare_3}): Large sit-and-stare with exposure time of $\approx$~8~s and cadence of $\approx$~9~s with a spectral binning of 2 for the FUV channel. We analyse about 20 minutes of the sit-and-stare observation from a start time of $\approx$~17:25UT.
    \item \textbf{Flare 4}: 2014-10-27 X-class flare (FL4, Figure~\ref{Fig:flare_4}): Large and coarse 8-step raster with raster cadence of $\approx$~26~s, exposure time of 2~s (down to $\approx$~0.26~s and 0.64~s for the FUV and NUV lines respectively) and step cadence of $\approx$~3~s. The observation has a spatial and spectral (for both the FUV and NUV channels) binning of 2 and a rotation angle of 90 degrees. We analyse several rasters covering about 18 minutes around the flare's peak time, from $\approx$~14:14 to 14:32UT. 
\end{itemize}

Figures \ref{Fig:flare_1}--\ref{Fig:flare_4} show an overview of FL1--4 respectively. For each figure, we are showing an overview of the flare as observed by the \iris~SJI 1400\AA~(dominated by \siiv~emission at T~$\approx$~80~kK) or 1330\AA~(dominated by \cii~emission at T$\approx$~10--40~kK) filters (panels a), and the AIA~131\AA~filter (which is dominated by \fexxi~emission at around 10~MK, panels g). Panels (b--e) and (h--m) show the spectrograph raster data within the field-of-view (FOV) which is overlaid on the SJI image in panels (a). For FL3, Figure~\ref{Fig:flare_3} shows the data across a portion of the slit (which is highlighted by two small horizontal marks in panel (a) as a function of time. FL2 and FL4 are rotated by 45 and 90 degrees respectively, but the raster data shown in Figures~\ref{Fig:flare_2} and \ref{Fig:flare_4} are not rotated for convenience. The spectrograph data in panels (b--e) and (h--m) show: \fexxi~intensity (b) and velocity (c), \oiv~intensity (d) obtained by performing single Gaussian fits in each pixel; \mgii~ k3 intensity (e) and velocity (h), k$_2$ peak difference ``\emph{diff}", i), depth of central reversal from the blue peak (i.e. ``\emph{dep}", l) and k$_2$ peak separation (``\emph{sep}", m). We defined \emph{diff}, \emph{dep} and \emph{sep} using the following formulae: 
\begin{equation}
\label{eq:diff}
   diff = \frac{I_{k2,R}-I_{k2,V}}{I_{k2,R}+I_{k2,V}}
\end{equation}

\begin{equation}
\label{eq:dep}
dep = - \frac{I_{k2,V}-I_{k3}}{I_{k2,V}+I_{k3}}
\end{equation}

\begin{equation}
\label{eq:sep}
sep = v_{k2,R}-v_{k2,V}
\end{equation}

where $I/v_{k3}$, $I/v_{k2,V}$ and $I/v_{k2,R}$ are the intensity and Doppler velocities of the line core, blue and red peaks of the \mgii~k line respectively (see Figure~\ref{Fig:profiles}). For each dataset, we obtained these values using the \texttt{iris\_get\_mg\_features\_lev2} routine available within the \iris\ solar software (SSW) package and described in \citet{Pereira2013}. Although this is an automatic method to detect the \mgii~line peaks, we also verified manually that it succeeded in fitting the RE profiles satisfactorily, in particular the line red and blue  peaks and the line core.  

The green contours in Figures~\ref{Fig:flare_1}--\ref{Fig:flare_4} represent the location of the \mgii~RE profiles that satisfy the criteria that we describe in Sect.~\ref{Sect:criteria}. Finally, the red continuous spectra shown in panels (f) and (n) were obtained by averaging the spectra in the pixels indicated by the green contours for the \fexxi~and \mgii~spectral windows respectively, while the grey dotted spectra indicate the 1-$\sigma$ standard deviation of the averaged spectra.
\subsection{Method to identify and characterize ribbon front profiles}\label{Sect:criteria}
After calculating the spectral parameters for the \mgii~line profiles as defined above, we identified the RE profiles by using the following metrics:

\begin{equation} 
diff \gtrsim 0
\end{equation}
\begin{equation} 
I_{k2,V} \gtrsim 0.8~I_{k3}
\end{equation}
\begin{equation} 
sep \gtrsim 30~km~s^{-1}
\end{equation}
\begin{equation} 
I_{m,t}  \gtrsim~20\%~I_{m,k}
\end{equation}
where $I_{m,t}$ and $I_{m,k}$ represent the maximum emission in the \mgii~triplet and k~line spectra respectively.
Such metrics were defined based on: 
\begin{enumerate}
\item the values we found in the RE profiles representative group by performing a $k$-means analysis for one of the flares (FL2) analysed in this work.
\item the range of values found in Figure~7 of \citet{Panos2021b}, who used machine learning techniques to obtain the most probable RE profiles for dozens of \iris~flares.
\end{enumerate}
    While a thorough machine learning analysis for the four flares analysed in this work is outside the scope of this paper (as it has been already presented in the series of papers by Panos et al.), method (1) allows us to put the RE metrics on a more quantitative basis, complementing the more qualitative overview presented in \cite{Xu2016} and \cite{Panos2018}.
Nevertheless, we verified that the two methods held  good agreement and we used a conservative approach when defining the metrics above, to ensure that the range of most probable RE profiles in Figure 7 of \citet{Panos2021b} would be included by our metrics and that we are not losing important information on RE profiles.

\begin{figure*}
\center
\includegraphics[width=\textwidth]{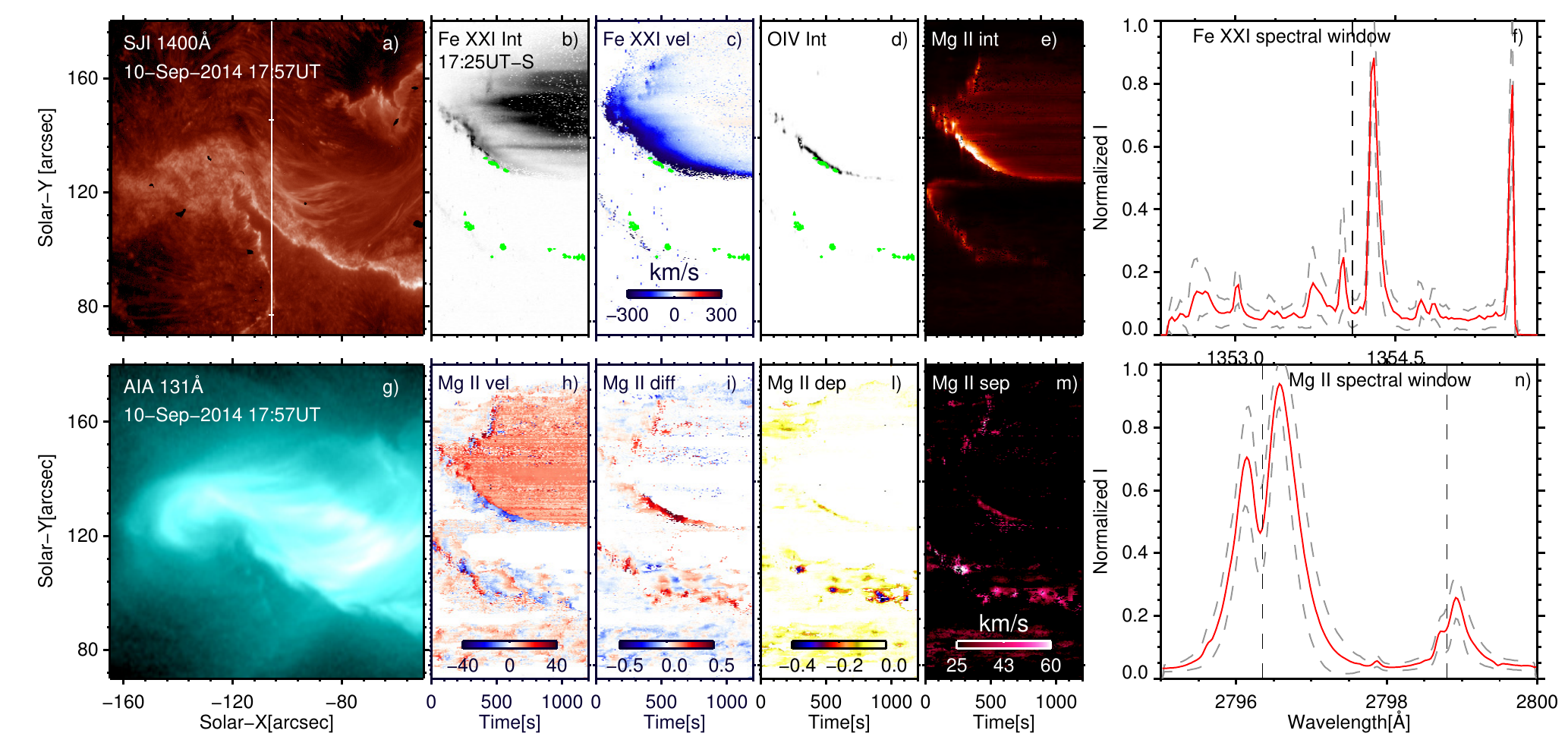}
\caption{Overview of FL3: for the panels' description, see Fig.~\ref{Fig:flare_1} and Sect.~\ref{Sect:obs}. The time in panel b) refers to the start time of the sit-and-stare observation shown here. The time in the x-axis for panels b)--e) and h)--m) is in seconds after the start time. Panels b)--e) and h)--m) show a portion of the sit-and-stare across the slit which is highlighted by two small horizontal marks on the slit in panel a).}
\label{Fig:flare_3}
\end{figure*}

\begin{figure*}
\center
\includegraphics[width=\textwidth]{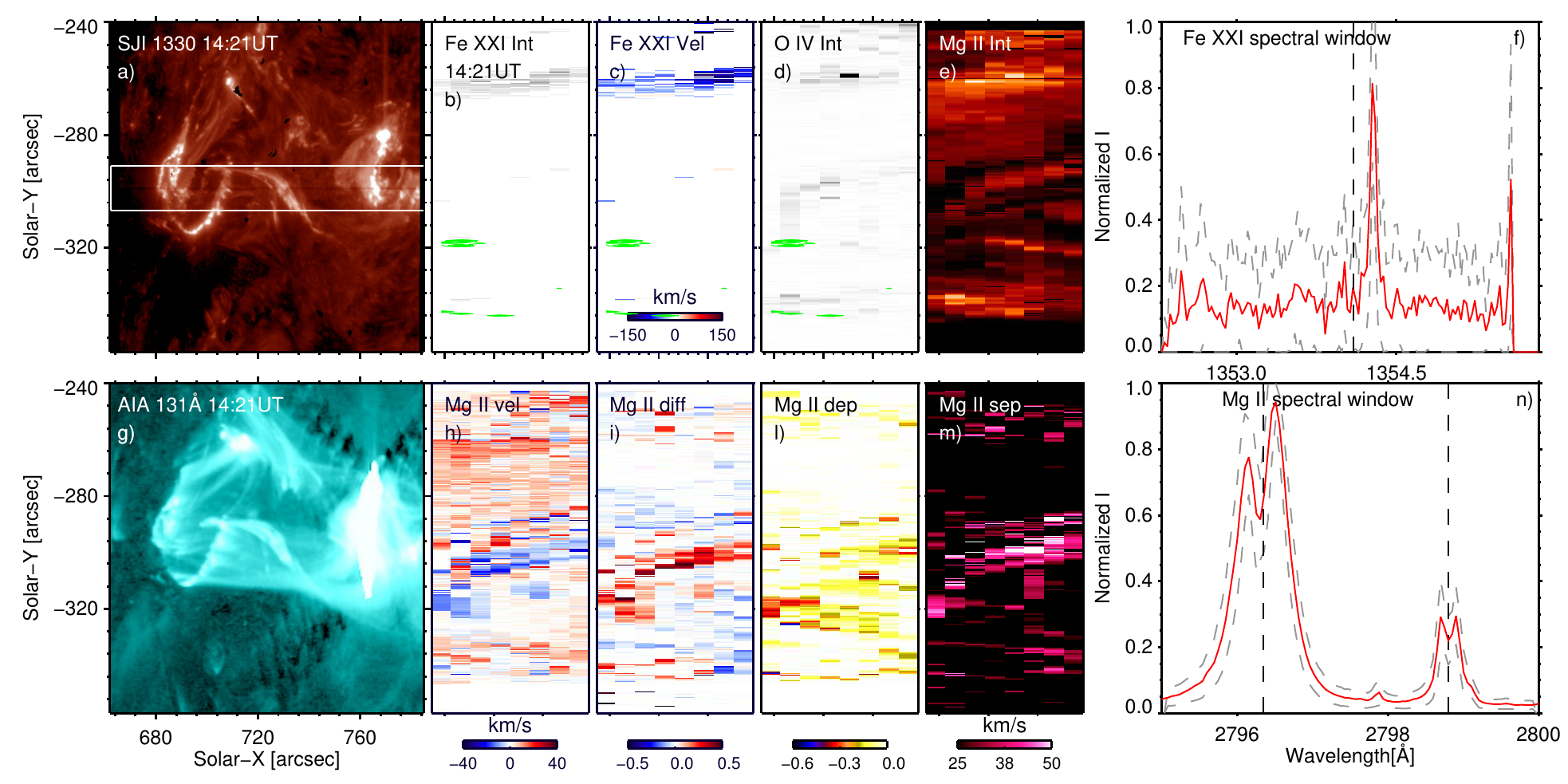}
\caption{Overview of FL4 (one example raster): for the panels' description, see Fig~ \ref{Fig:flare_1} and Sect.~\ref{Sect:obs}. This observation had a rotation angle of 90 degrees, but the raster data in panels b)--e) and h)--m) are not rotated for convenience. The time in panel b) refers to the mid-time of the example raster (the raster cadence is $\approx$~26s). A movie associated with this figure is also available. } 
\label{Fig:flare_4}
\end{figure*}

We note that our analysis does not distinguish between different groups of RE profiles that are discussed in Panos' analysis, e.g. groups 11, 12 and 52 in Figure 3 of \citet{Panos2018}, as well as ``strong" and ``weak" RE profiles in Figure 7 of \citet{Panos2021a}. This means that our averaged profiles in Figures~\ref{Fig:flare_1}--\ref{Fig:flare_4} might show less pronounced features (in particular depth of the central reversal and peak separation) than those in group 52 or the ``strong" RE profiles presented in those papers. We also note that the more extreme profiles (or group 52) are not as common as the weaker RE profiles (or groups 11-12), as can be seen in Figure 8 of \cite{Panos2018}. In particular, flares 22, 29 and 31 in that figure correspond to our flares FL2, 3 and 4 respectively. We note that our FL3 has the lowest incidence of strong RE profiles, while FL2 has the largest. Even then, the ``weaker" RE profiles (closer to our average profiles in Figures~\ref{Fig:flare_1}--\ref{Fig:flare_4}) dominate.

Figures \ref{Fig:flare_1}--\ref{Fig:flare_4} show that the profiles identified by these criteria are indeed found in the ribbon leading edge. In addition, we note that \fexxi~and \oiv~TR emissions are fainter or not visible in the RE locations. The fact that there is little or no \fexxi~evaporation is in agreement with what is shown in Figure 7 of \citet{Panos2021b}. In this figure the authors show that the most probable spectra (red) associated with the RE profiles (top panels) do not contain clear \fexxi~emission, or even when they do (less probable spectra in blue), the line is often not significantly blueshifted, or in other words there is \emph{no clear chromospheric evaporation simultaneous with RE profiles}. On the other hand, the evaporation is stronger in the trailing front (middle and bottom panels of that figure). As also mentioned in \citet{Panos2021b}, in principle this could be due to a delay in the formation of the \fexxi~line. The FL3 sit-and-stare observation we analyse here is best suited to investigate the time delay between the appearance of the RE profiles and evaporation, thanks to higher cadence (around 9s). Panels (c) and (d) of Figure~\ref{Fig:flare_3} show that the ``southern" branch of the single ribbon  observed under the \iris~slit shows a number of pixels where the RE \mgii~profiles have been identified, but no \fexxi~emission or evaporation is visible there at any later time. We note that this might be partly caused by the high inclination of the flare loops. For the few pixels in the ``upper" branch where do we see evaporation at a later time (green contours in Panels b and c), we calculated an average delay of 45~s between the appearance of the RE profiles and the \fexxi~evaporation. 

Finally, we summarize the observational parameters of the RE  \mgii~profiles for FL1--4 in Figures~\ref{Fig:hist_gradual} and \ref{Fig:hist_const} and Table~\ref{table_sims_iris}, as will be discussed in the following sections. 


\section{Forward Modelling of IRIS Flare Emission}
\label{Sect:sims}
In this section we describe the method we use to simulate the \iris~synthetic spectra using a grid of \radynfp~simulations (Sect.~\ref{Sect:radyn}). We post-process \radynfp\ atmospheres through the \rhtiago~code for the synthesis of the \mgii\ emission (Sect.~\ref{Sect:RH}) and the CHIANTI v.10 \citep{Dere1997,DelZanna2021} atomic database for the optically thin emission (Sect.~\ref{Sect:thin}). The synthethic spectra and parameters that we obtain from the simulations are shown in Figures~\ref{Fig:hist_gradual}, ~\ref{Fig:hist_const} and \ref{Fig:summary_sims} and in Figures~\ref{Fig:mgii_gradual}--\ref{Fig:mgii_tripl_const} in the Appendix, summarized in Table~\ref{table_sims_iris} and finally discussed in detail in Sect.~\ref{Sect:discussion}.
\subsection{RADYN simulations}
\label{Sect:radyn}
To investigate the origin of the enhanced absorption of the \hei~10830~\AA\ line at the flare leading edge, \cite{2021ApJ...912..153K} produced a large grid of flare simulations using the \radyn\ radiation hydrodynamic model \citep{1995ApJ...440L..29C,2005ApJ...630..573A,2015ApJ...809..104A}, which uses the non-thermal particle transport code \fpcode\ \citep{2020ApJ...902...16A} to propagate a distribution of non-thermal particles through a flare loop. We use a subset of those simulations here, synthesising from them the \mgii\ NUV spectra as well as \fexxi\ and \oiv\ emission. For full details of these \radynfp\ simulations, and the code in general, consult \cite{2021ApJ...912..153K}. 

The simulations selected for use here covered a wide range of energy fluxes, with energy deposition via an injected distribution of non-thermal electrons. The fluence (time-integrated energy flux) was varied, with $F = [1\times10^{10}, 5\times10^{10}, 1\times10^{11}, 5\times10^{11}, 1\times10^{12}]$~erg~cm$^{-2}$. These electron beams were injected for either $t_{inj} = 10$~s (a constant flux) or $t_{inj} = 20$~s (a triangular profile with a peak at $10$~s), the latter to investigate a more gradual injection. In the remainder of the text we refer to the instantaneous energy flux (erg~cm$^{-2}$~s$^{-1}$) alongside the injection time, as this property is more commonly used in the flare literature, where for the $t_{inj}=20$~s cases we quote the peak instantaneous energy flux.
The spectral shape of those distributions was varied also. The spectral index $\delta = 5$ was fixed but two values of the distribution's low-energy cutoff were studied $E_{c} = [15,30]$~keV, allowing us to study the difference between a `softer' or `harder' non-thermal electron spectrum. The latter contains a larger proportion of higher energy electrons, capable of penetrating more deeply and resulting in a smaller amount of heating of the upper chromosphere/lower transition region. This was motivated because \cite{2021ApJ...912..153K} found that a harder distribution, with a weaker energy flux, resulted in stronger, slightly longer-lived periods of enhanced absorption of the \hei~10830~\AA\ line \citep[i.e. those simulations were more consistent with the ribbon front observation of][]{Xu2016}. We now ask ``Do the synthetic \mgii\ line profiles for those same simulations similarly appear more consistent with the observed ribbon front profiles?'' 

For comparison, we also analyse a flare simulation that is more efficient at heating the transition region and driving chromospheric evaporation. That is, a simulation with a large impulsive energy flux (10$^{11}$~erg~cm$^{-2}$~s$^{-1}$ with constant $t_{inj}$ = 10s), low-energy cut-off ($E_{c}$ = 10keV) and $\delta =5$. 
\subsection{Synthesis of \mgii~emission}
\label{Sect:RH}
To synthesise the \mgii\ NUV spectra from our \radynfp\ simulations we used the radiation transport code \rhtiago\ \citep{2001ApJ...557..389U}, which solves the equation of radiation transport and atomic level populations given an input atmosphere. Flare atmospheres (temperature, electron density, bulk velocity, hydrogen atomic level populations) were input to \rhtiago, with a cadence of 0.5~s. The NLTE radiation transport was solved for H, \mgii and \caii, with 15 additional species solved in LTE as sources of background opacity. An additional source of line broadening due to microturbulence was included, with a constant value of $7$~km~s$^{-1}$ \citep[consistent with][]{Carlsson2015}. The atmosphere above a temperature of 30~kK was discarded in the solution to reduce computational time.

When solving the \mgii\ radiation transfer we used the 10 level-plus-continuum model atom from \cite{2013ApJ...772...89L}, and included the effects of partial frequency redistribution via the hybrid scheme of \cite{2012A&A...543A.109L}, which has been shown to be required in both quiet Sun and in flares \citep{2013ApJ...772...89L, 2019ApJ...883...57K}. While \radynfp\ includes non-equilibrium ionisation, \rhtiago\ does not, solving each timestep in isolation assuming statistical equilibrium. This is somewhat mitigated by using the non-equilibrium electron density from \radynfp\, but it was also demonstrated that the assumption of statistical equilibrium is largely sufficient for \mgii\ even in flares \citep{2019ApJ...885..119K}.

The synthetic \mgii\ NUV spectra were converted to the IRIS count rates by: (1) convolution with a spectral PSF assumed to be a Gaussian with FWHM of $52$~mA (two IRIS spectral pixels), (2) recasting to the IRIS spectral plate scale ($26$~mA) and multiplying by the spectral dispersion, (3) converting intensity from ergs to photons, (4) multiplying by the solid angle subtended by an IRIS SG pixel, (5) multiplying by the IRIS effective area, and (6) converting from photons s$^{-1}$ to DN~s$^{-1}$ \citep[18 photons DN$^{-1}$ in the NUV, ][]{DePontieu2014}. An exposure time of 1~s was assumed. Finally, the same metrics as calculated for the observtions were calculated from these synthetic IRIS spectra (see Sect.~\ref{Sect:criteria}).

In the Appendix, we show the synthetic spectra of \mgii~and \mgii~triplet as a function of time for the subset of the models from \citet{2021ApJ...912..153K} that we analyse here. 
\subsection{Synthesis of optically thin \fexxi~and \oiv~emission}
\label{Sect:thin}
Similarly to what was done in our previous work \citep{Polito2018, Polito2019}, we synthesize the emission of the optically thin \fexxi~and \oiv~lines using the values of density, temperature, and bulk velocity at each grid point and timestep from the \radynfp~ simulations and atomic data from CHIANTI v.10 \citep{Dere1997,DelZanna2021} assuming photosperic abundances \citep{Asplund2009}. In particular, we follow Eq. 1 of  \citet{Polito2018} and convert the synthethic spectra in units of DN s$^{-1}$ pixel$^{-1}$ assuming the \iris~unsummed spatial pixel dimension (e.g. 0.33\arcsec $\cdot$ 0.166\arcsec), a spectral bin of 26~m\AA\ and a gain of 4 photons DN$^{-1}$ for the FUV channel \citep{DePontieu2014}. The time-velocity spectra in Figure~\ref{Fig:summary_sims} are then obtained by integrating the synthetic emission in each \radynfp~ grid points along the loop as a function of time, and are plotted every 1s. Finally, we take into account the \iris~instrumental broadening of 26~m\AA\ when synthesizing the line emission.

\begin{figure*}[!ht]
\center
\includegraphics[width=\textwidth]{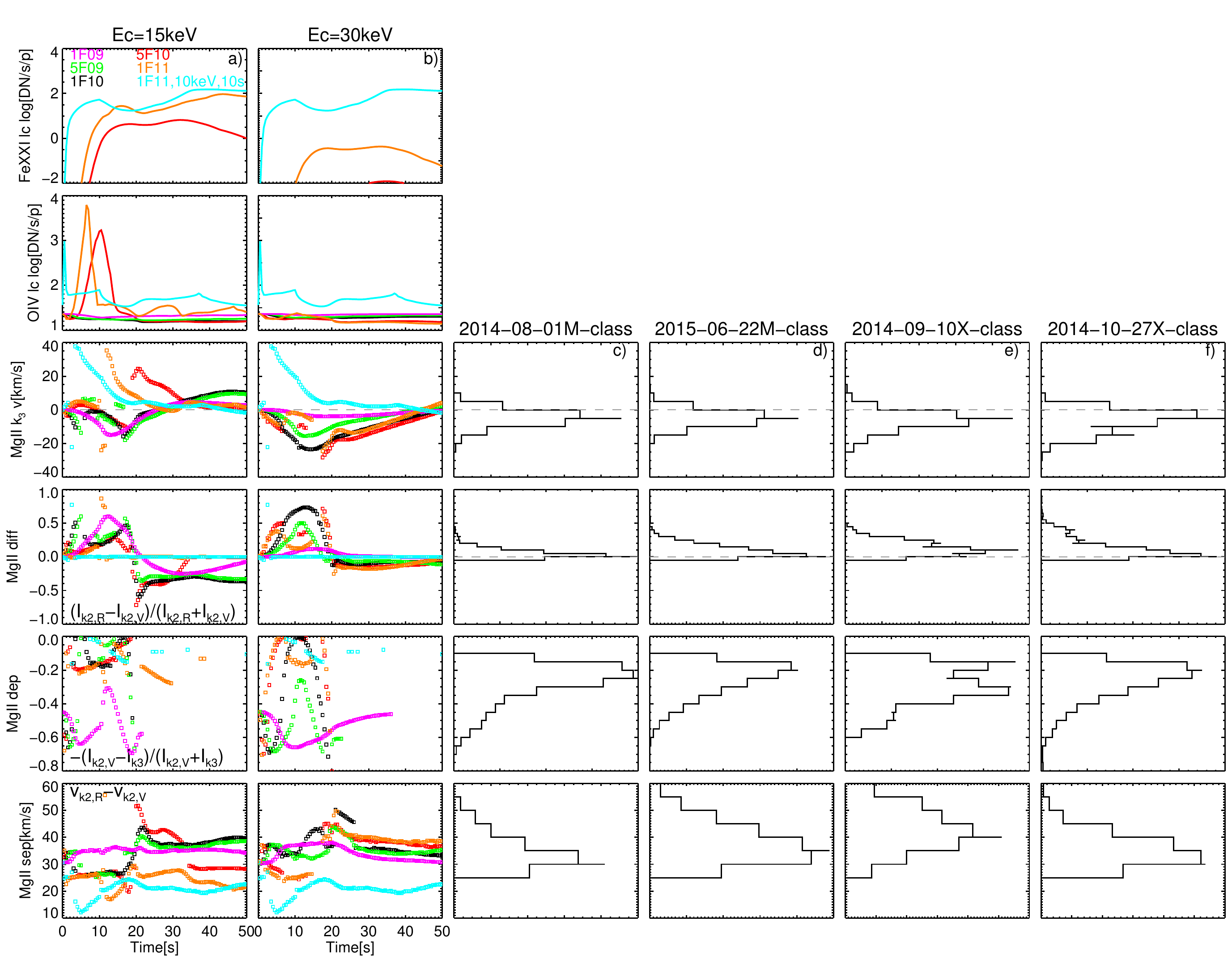}
\caption{Columns a--b: summary of parameters from \radynfp+\rhtiago~simulations assuming \textit{gradual} heating over 20s as a function of time. Gaps in the synthetic curves with square symbols indicate time steps in the simulations where the line goes single peaked, triple peaked or where the fit fails. Columns c--f: comparison with histograms of observed RE metrics for Flares 1--4.  }
\label{Fig:hist_gradual}
\end{figure*}
\begin{figure*}[ht]
\center
\includegraphics[width=\textwidth]{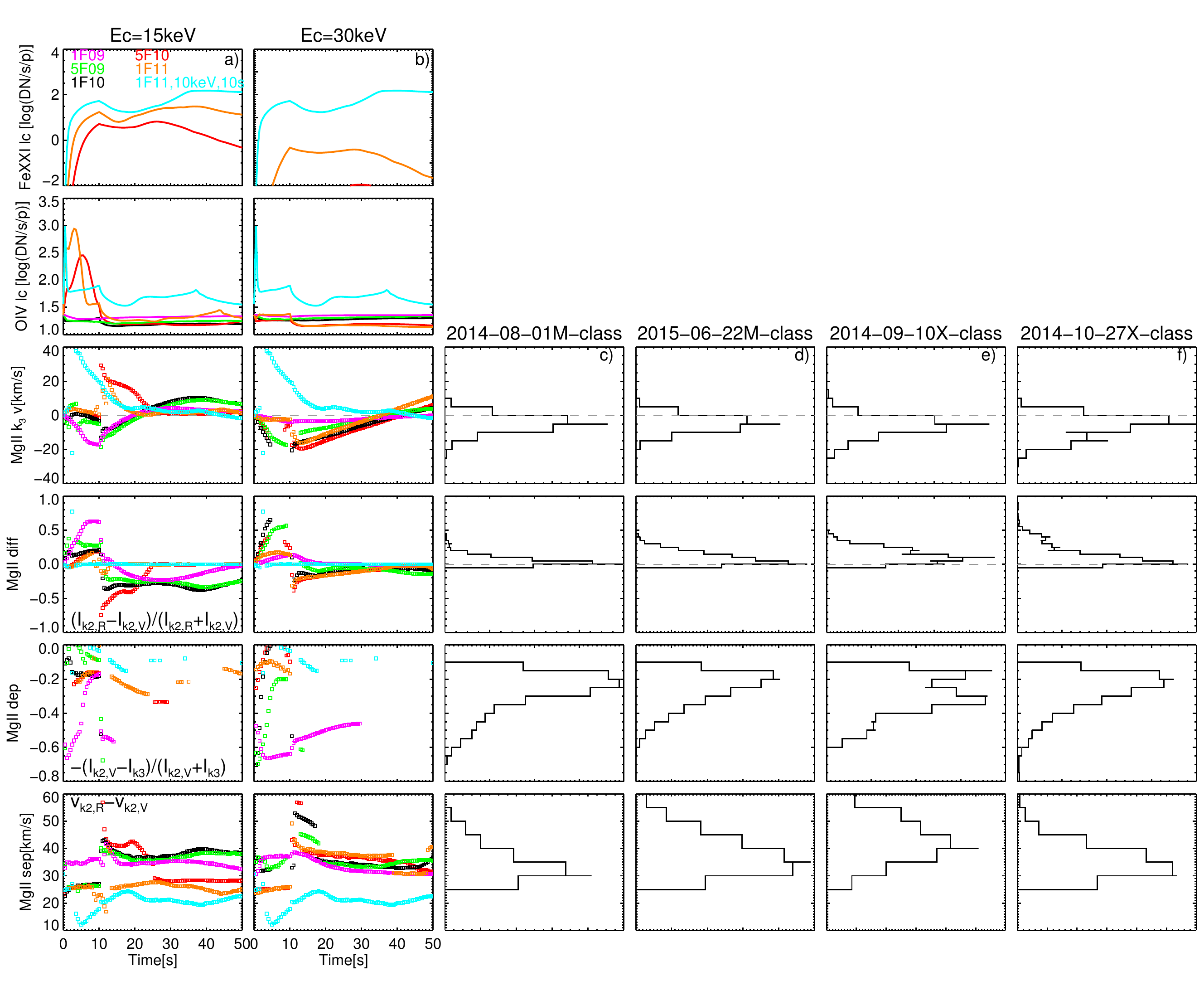}
\caption{Same as Fig~\ref{Fig:hist_gradual} for models with constant 10s heating.
} 
\label{Fig:hist_const}
\end{figure*}
\begin{figure*}
\center
\includegraphics[width=\textwidth]{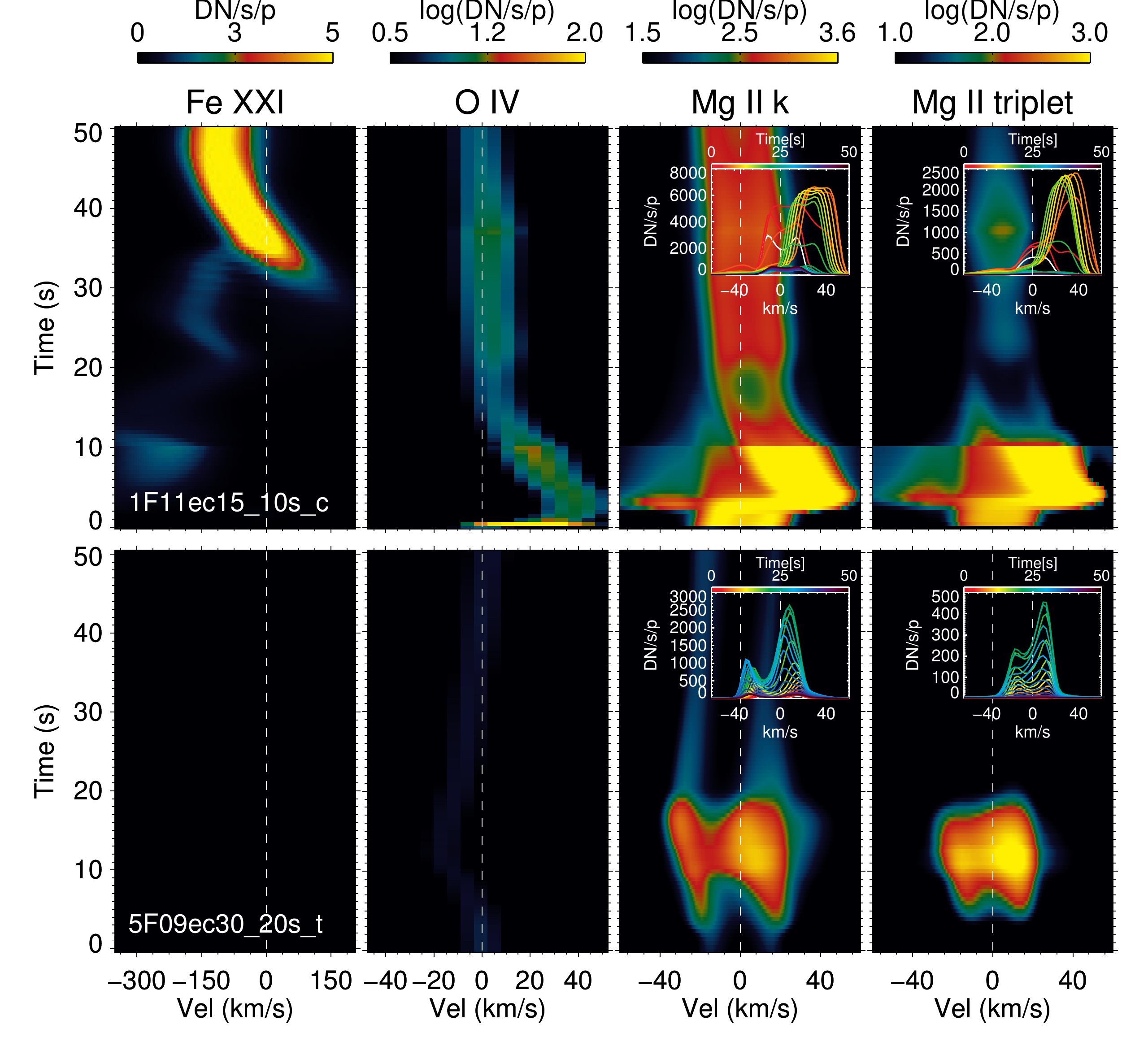}
\caption{From left to right:  \fexxi, \oiv, \mgii~k and triplet synthetic spectra for a ``typical" evaporation model (top) and a model that we suggest reproduce quantitatively the ribbon leading edge behaviours (bottom). The inserts in the \mgii~k and triplet spectra show the evolution of the synthetic spectra ever 4s. For a summary of all the models results, see Tab.~\ref{table_sims_iris} and Figures~\ref{Fig:mgii_gradual}--\ref{Fig:mgii_tripl_const} in the Appendix.} 
\label{Fig:summary_sims}
\end{figure*}

\begin{deluxetable*}{cccccccccc}[!h]
\tablecaption{Spectral characteristics of IRIS lines for different \radynfp~models}
\tablehead{
 &  \colhead{Model}   & & & & & \colhead{IRIS Spectral features} & & &  \\ \cline{1-3}  \cline{5-10}
  \colhead{F[erg/cm$^{2}$/s]} & \colhead{E$_C$[keV]} & \colhead{$\tau_{H}$[s]}&  & \colhead{\fexxi~em\tablenotemark{a}}&\colhead{\oiv~ em\tablenotemark{b}}& \colhead{$\nu_{\oiv}$[km/s]\tablenotemark{c}}&\colhead{DCR\tablenotemark{d}} &  \colhead{$\nu_{CR}$[km/s]\tablenotemark{e}} &\colhead{tDCR\tablenotemark{f}} 
}
\startdata
10$^9$ & 15 & 20(t) & & No & No & Blue & Yes & Blue & No \\
10$^9$ & 30 & 20(t) & & No & No & Bluewing &Yes & Blue& Yes \\
5 $\cdot$ 10$^9$ & 15 & 20(t) & & No & No &Blue& No&Blue & No\\
5 $\cdot$ 10$^9$ & 30 & 20(t) & & No& No&Blue &Yes&Blue & Yes\\
10$^{10}$ & 15 & 20(t) & & No& No & Blue&No &Blue& No\\
10$^{10}$ & 30 & 20(t) & & No& No &Blue &Yes &Blue & Yes\\
5 $\cdot$ 10$^{10}$ & 15 & 20(t) & & Yes&Yes &Blue& No &Blue& No\\
5 $\cdot$ 10$^{10}$ & 30 & 20(t) & &No & No&Blue &No& Blue& No\\
10$^{11}$ & 15 & 20(t) & & Yes& Yes& Blue/Red(blue followed by red)&No&Red &No \\
10$^{11}$ & 30 & 20(t) & &No &No &Blue&Yes &Blue& No\\
10$^9$ & 15 & 10(c) & &No & No&Blue &Yes& Blue& No\\
10$^9$ & 30 & 10(c) & &No & No& Blue&Yes& Blue& Yes\\
5 $\cdot$ 10$^9$ & 15 & 10(c) & & No& No&Blue& No& Blue& No\\
5 $\cdot$ 10$^9$ & 30 & 10(c) & &No & No&Blue& No&Blue & No\\
10$^{10}$ & 15 & 10(c) & &Yes & Yes&Blue &No &Red& No\\
10$^{10}$ & 30 & 10(c) & & No& No&Blue& No& Blue& No\\
5 $\cdot$ 10$^{10}$ & 15 & 10(c) & & Yes& Yes& Blue/Red&Yes & Red& No\\
5 $\cdot$ 10$^{10}$ & 30 & 10(c) & & No& No&Blue& Yes& Blue& No\\
10$^{11}$ & 15 & 10(c) & &Yes& Yes&Blue/Red &Yes& Red& No\\
10$^{11}$ & 30 & 10(c) & &No& No& Blue&Yes& Blue& No\\
10$^{11}$\tablenotemark{*} & 10 & 10(c) & &Yes &Yes&Red& No& Red& No\\
 \textbf{Observation of RE profiles} & & &&No&No&Blue/Red\tablenotemark{**} &Yes&Blue&Yes/No\\
\enddata
\label{table_sims_iris}
\tablenotetext{a}{Significant \fexxi~emission}
\tablenotetext{b}{Increased \oiv~emission}
\tablenotetext{c}{``Deep" Central Reversal in \mgii~k3}
\tablenotetext{d}{Doppler shift of \mgii~k3 central reversal in km/s}
\tablenotetext{e}{``Deep" central reversal in \mgii~triplet}
\tablenotetext{*}{``Evaporation" simulation}
\tablenotetext{**}{See Sect.~\ref{Sect:TR}.}
\end{deluxetable*}
\section{Discussion}
\label{Sect:discussion}
\subsection{Model-data comparisons}
In Figures~\ref{Fig:hist_gradual} and \ref{Fig:hist_const} we compare the observational results from \iris~with the predictions from the \radynfp\ and \rhtiago\ models, with gradual and constant heating profiles respectively. For each figure, the first two columns from the left show the model predictions for models with E$_C$ of 15 and 30~keV respectively. From top to bottom, we plot the: \fexxi~and \oiv~intensities, \mgii~k$_3$ line core velocity, peak difference, depth of central reversal and peak separation as a function of time. To calculate and define the \mgii~spectral parameters we have used the same method and  definitions (e.g. Eqs.~\ref{eq:diff}--\ref{eq:sep}) as those used in Sect.~\ref{Sect:obs} for the observations. The third to sixth columns in Figures~\ref{Fig:hist_gradual} and \ref{Fig:hist_const} show histograms of the same \mgii~parameters that we have obtained for the four flare observations described in Sect.~\ref{Sect:obs}. We note that we do not report the values of \fexxi~and \oiv~intensities for the observations since these lines are often not observed or very faint in the same pixels where we see the RE profiles. In addition to the \radynfp\ models presented in \cite{2021ApJ...912..153K}, the synthetic parameters for a more ``typical'' flare simulation that is efficient at driving evaporation (see Sect.~\ref{Sect:radyn}) are also shown in light blue color in Figures~\ref{Fig:hist_gradual} and \ref{Fig:hist_const}.

Table~\ref{table_sims_iris} also summarises the quantitative behaviour of the \iris~spectral lines in the models analysed here as well as the observations. The entries in the observational row are taken from the analysis of the datasets presented here which are also consistent qualitatively with  \citet{Panos2018,Panos2021a}.

Below we summarize the main findings from our model-data comparison based on Figures~\ref{Fig:hist_gradual} and \ref{Fig:hist_const} and Table~\ref{table_sims_iris}: 
\begin{itemize}
    \item The ``typical" flare simulation with large impulsive flux (10$^{11}$~erg~cm$^{-2}$~s$^{-1}$ with constant $t_{inj}=10$~s) and  $E_{c}=10$~keV produces the strongest \fexxi~and TR emission. This is not surprising since the strong energy flux with a softer low energy cut-off means that most of the energy is deposited in the TR where it quickly drives the plasma to million degrees temperatures where \fexxi~is formed. When the plasma is not able to radiate the energy away, the overpressure will drive the chromospheric evaporation \citep[e.g.][]{1985ApJ...289..414F}. However, the same simulation does not seem to reproduce the more typical properties of the RE profiles. In particular, the synthethic profiles show that the core of the \mgii~lines are mostly redshifted, with shallow central reversal and small peak separation.
    
    \item Most of the electron beam models above (from \citet{2021ApJ...912..153K}), apart from the ``evaporation" model (light blue curves) can reproduce, even if just for a short time, asymmetric profiles with stronger red peak and slightly blueshifted line core, consistent with \iris\ observations. Also, the range of blueshifts for the line core seems to reproduce the magnitude of those seen in the observations. The values of peak separation in the models from \citet{2021ApJ...912..153K} also reproduce the observed values. This might be due to the fact that we added a microturbulence of $7$~km~s$^{-1}$ in the \rhtiago\ models, following \citet{Polito2018,Carlsson2015}. 
    
    \item The spectral parameter that best distinguishes the models is the \emph{depth of the central reversal}. We emphasize that in the observations we have made no distinction between the two types of ``weak" and ``strong" ribbon profiles of \citet{Panos2021a}, and that the strong tail of deeper central reversal values in the histograms of Figures~\ref{Fig:hist_gradual} and ~\ref{Fig:hist_const} are more representative of the strongest profiles with deeper reversal. For the gradual heating models (Figure~\ref{Fig:hist_gradual}), the simulations that reproduce the strongest central reversals in \mgii~are those with the more modest energy fluxes of 1--5F9 (magenta color) for the smaller E$_C$ (first column). However, even gradual heating models with higher energies fluxes up to 1F11 can to some extent explain the deep central reversals if the E$_C$ is larger (30~keV). For a fixed flux, electron beams with stronger E$_C$ contain a larger fraction of high-energy electrons, capable of penetrating deeper into the atmosphere in the formation region of the line peaks. There the beam heating will drive an increase in the plasma density that causes the stronger reversal of the line core compared to the peaks.
    
    \item The constant heating models are less successful at explaining the strongest deep central reversals. In particular, the ones that work best are again the ones with more modest (1--5F09) energy flux and larger E$_C$. For the smaller E$_C$ of 15 keV (softer beams) only the 1F09 model can explain the very large central reversals of the ``strong" ribbon front profiles. 
    
    \item The simulations that can reproduce a deep central reversal for the longest time are the gradual heating simulations with 1--5F9 with E$_C=30$~keV, the 1F9 simulation with E$_C=15$~keV, and the 1F9 constant heating simulation with E$_C=30$~keV. However, the 1F9 simulations deposit very little energy in the atmosphere and produce faint line emission for both the \mgii~and \mgii~triplet. In the observations the \mgii~line is often observed to be fainter at the ribbon front profiles (see Figures~\ref{Fig:flare_1}--\ref{Fig:flare_4}) and also the \mgii~triplet emission can vary based on the observation.

\end{itemize}

 In addition to what is discussed above, in Table~\ref{table_sims_iris} we also add the information regarding the TR Doppler shift for all models. One thing that seems to be discrepant between models and observations in same cases is the Doppler shift of the TR lines. This topic is discussed separately in Sect.~\ref{Sect:TR}.
 
To summarize, the model-data comparison above seem to suggest that we need two types of substantially different models to explain the behaviour of the ``trailing edge" of the ribbon where the evaporation is observed and the ``leading edge" of the ribbon where the typical profiles identified by \citet{Xu2016} and the series of papers by Panos et al. 

In Figure~\ref{Fig:summary_sims}
 we show the synthetic spectra over time for \fexxi, \oiv, \mgii~k~and \mgii~triplet for a more ``typical evaporation" simulation (top panels), which we speculate could be representative of the heating mechanism for the ribbon trailing edge, and one of the gradual heating simulations that best reproduces quantitatively the behavior of the ribbon front profiles observed with \iris~(bottom panel). We also note that this type of simulation (gradual and modest energy release and harder beam) can also reproduce the increased absorption in \hei 10830~\AA\ that has been observed by \citet{Xu2016}, as shown in \citet{2021ApJ...912..153K}. 

However, it is important to note that while the ``typical evaporation" simulation produces \mgii\ h \& k lines with much shallower reversals, it still cannot reproduce the single-peaked behaviour of the \mgii~line typically seen in the trailing edge of the ribbon. This is a common problem that has been discussed by several authors, that seems to be the result of an underestimation of electron density in the upper chromosphere \citep[e.g.][]{2017ApJ...842...82R,2019ApJ...879...19Z}.

As mentioned in Sect.~\ref{Sect:obs}, in our FL2 observation we see a delay of about 45~s between the appearance of the RE profiles and the \fexxi~evaporation. In addition, \citet{Panos2021a} states that on average it takes about 1--3 minutes for the RE profiles to become single peaked. Since our loops are heated for 20s, one might wonder whether extending the duration of the heating in one of the simulations that reproduces the ribbon front profiles might naturally lead to \fexxi~evaporation. While we are working on a follow-up paper focused on the long duration of the ribbon front profiles (Kerr et al, \textsl{In Prep.}), we have also performed preliminary experiments to explore this possibility, as discussed below (Sect.~\ref{Sect:extra}).

\subsection{Do we need different models for the trailing and leading edge heating?}
\label{Sect:extra}
Based on the comparison between models and observations presented above, we speculate that we need different types of electron beam models to explain heating in different parts of the ribbons at a certain time, namely the ``leading" and  ``trailing" edges. However, since in some cases the same  location where we see the ribbon front profiles later can show  the typical features of the ribbon ``trailing" edge (i.e. \fexxi~evaporation, increased TR emission and single peaked \mgii~profiles) an alternative explanation could be that the same heating models that initially reproduce the ribbon front profiles then naturally also produce these typical features. Since the delay between the two regimes of behaviours can be a few tens of seconds as discussed above, the simulations from \citet{2021ApJ...912..153K} that we have discussed so far cannot directly answer this question since they are assuming a heating duration of up to 20s and the total duration the simulations is to 50s. To address this issue we ran some additional \radynfp\ simulations where we used the same electron beam parameters as in the model that we have chosen to be representative of a ``ribbon front-type" of heating (Figure~\ref{Fig:summary_sims}, bottom panels) but with longer duration, as summarized below:

\begin{enumerate}
\item A model with a gradual triangle heating profile with peak energy flux of 1.67~$\cdot$~10$^9$ ergs cm$^{-2}$ s$^{-1}$ (1.67F09), E$_C= 30$~keV and $\delta=5$, where the heating is released over 60~s instead of 20~s. The total energy integrated over time will be the same as that of the the 5F09 simulation with E$_C = 30$~keV and $\delta=5$ that we have described in the previous sections. 
    \item A model with a gradual triangle heating profile with peak energy flux of 5~$\cdot$~10$^9$ ergs cm$^{-2}$ s$^{-1}$ (5F09), E$_C= 30$~keV and $\delta=5$, where however the heating is released over 60~s instead of 20~s. The total energy integrated over time will be higher than the 5F09 simulation with E$_C= 30$~keV and $\delta=5$ that we have described in the previous sections. 
    
    \item A model with a gradual rising phase that is the same as the 5F09 simulation with E$_C= 30$~keV that is shown in Figure~\ref{Fig:summary_sims} (i.e. that reaches a peak energy flux of 5~$\cdot$~10$^9$ ergs cm$^{-2}$ s$^{-1}$ at $t=10$~s) but it then stays constant for a further 110~s. 
\end{enumerate} 
We also let the 3 simulations evolve until they reach  120s. These simulations all reproduce \hei\ 10830~\AA\ enhanced absorption followed by emission (not shown here). In Figure~\ref{Fig:long_sims} in the Appendix we show the synthetic spectra of the \fexxi, \oiv, \mgii~and \mgii~triplet lines for the additional models described above. The spectral characteristics of the \mgii~k~and triplet lines are similar to those observed in the bottom panels of Figure~\ref{Fig:summary_sims}, but the longer heating duration does not naturally lead to \fexxi~evaporation and increased TR emission as more typically observed in the trailing edge. We also emphasize that the simulation that produces these latter behaviour (e.g. top panels of Figure~\ref{Fig:summary_sims}), does not reproduce the ribbon front typical profiles before the onset of the evaporation either. We then conclude that the heating models that drive these two different regimes must be different, or that the electron beam parameters change over time. One other point that was also discussed in \citet{2021ApJ...912..153K} is that both the enhanced absorption of \hei 10830~\AA\ and the \mgii~ribbon front profiles \citep[e.g.][]{Panos2021a} can be sometimes observed for a few minutes, while our models can reproduce these behaviours for a period closer to a few seconds at most. A follow-up work focused on the long term evolution of the ribbon fronts profiles is currently under preparation (Kerr et al, \textsl{In Prep.}). 
\begin{figure*}
\center
\includegraphics[width=0.7\textwidth]{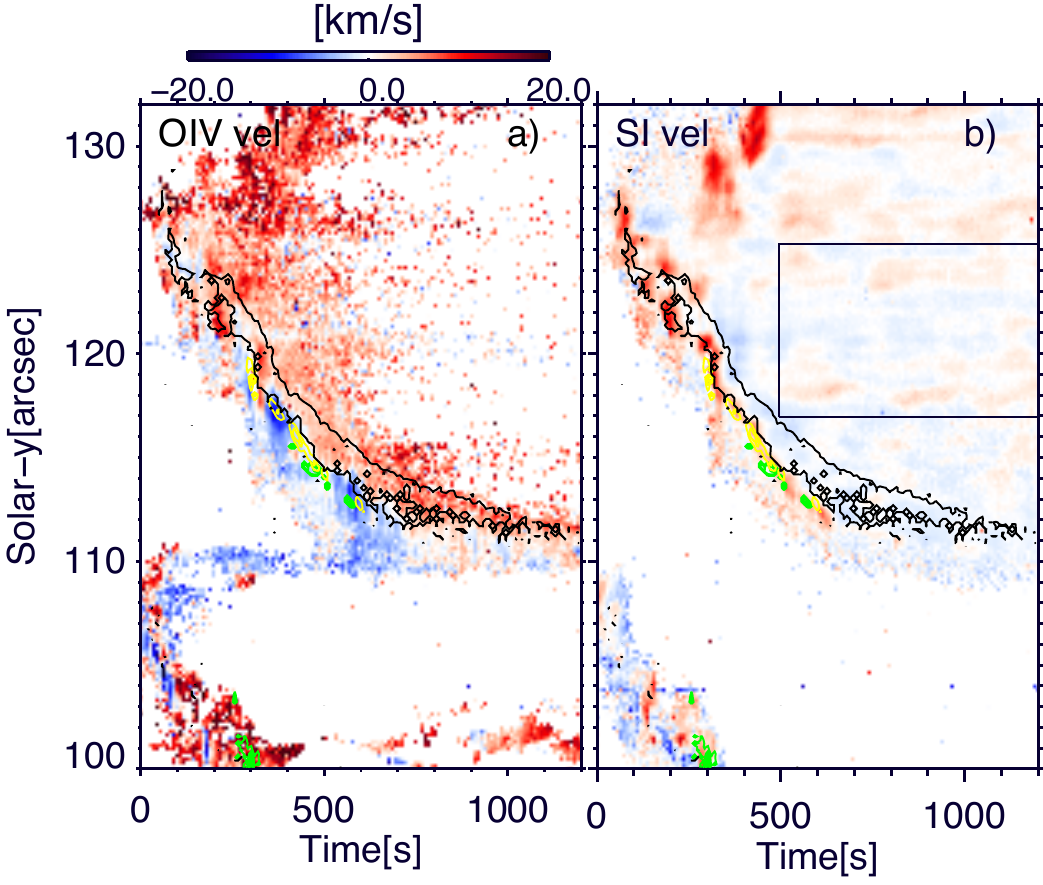}
\caption{Doppler shift maps of \oiv~and \si~around 1401\AA~for flare Fl 3. The green, yellow and black contours show the location of: the ribbon leading edge pixels, the maximum \oiv~intensity and the maximum \fexxi~evaporation. The box in the \si~map shows the area where we measured the average Doppler shift to verify wavelength calibration. See text for more details.} 
\label{Fig:doppler}
\end{figure*}

\subsection{Transition region emission}
\label{Sect:TR}
Flare ribbon observations most commonly show that the TR emission is redshifted in the ribbon area \citep[e.g.,][]{Tian2015,Polito2016,Reep2018}, with some exceptions \citep[e.g.,][]{Jeffreyeaav2794, Lorincik2022}. Nevertheless not many authors so far have focused on the local differences in the TR spectra between the ribbon trailing and leading edge locations.
We analysed in detail the \oiv~spectra in the ribbon front region for two of the four flares under study (see Figure~\ref{Fig:doppler} and \ref{Fig:doppler2}), and found that the line appears to be blueshifted at least in some locations within the ribbon fronts.  

Figure~\ref{Fig:doppler} shows  Doppler shift maps of the \oiv~and \si~line for FL3 as obtained by performing a single Gaussian fit in each \iris~pixel. The \iris~level 2 data are corrected for the orbital drift of the wavelength array. However, in order to perform an absolute wavelength calibration, it is usually recommended to measure the wavelength position of one of the photospheric lines included in the \iris~ spectra in case a small residual drift is still present. The closest photospheric line to the \oiv~lines is the \si~line around 1401~\AA. To verify the wavelength calibration, we took the average wavelength of the \si~line in a region along the flare loops during the gradual phase, and obtained a velocity of $\approx~0$~km s$^{-1}$, as expected if the wavelength calibration is correct. Note that the line is too faint to be observed reliably outside the flare.

The green, yellow and black contours in the Figure show the location of: the ribbon leading edge pixels, the maximum \oiv~intensity and the maximum \fexxi~evaporation respectively. The \oiv~Doppler map thus shows that in the ribbon leading edge the line is fainter (but still reliably measurable as we verified while performing the fit) and blueshifted. The line is instead redshifted in the trailing ribbon where the evaporation (black contours) is observed.

Similarly, Figure~\ref{Fig:doppler2} shows the \oiv~Doppler shifts for two example rasters during FL2 (a movie is also available). Since the signal for this flare was sometimes small given the short exposure time ($\sim1$~s), we binned the data by an additional factor of 2 in the Y-direction. The Figure and corresponding movie suggest that the \oiv~Doppler shift is sometimes blueshifted and other times redshifted in the location of the ribbon front profiles. 
 
We also note that \citet{Panos2021b} showed the characteristics \siiv~profiles in the ribbon front location in Figure 7 of their paper. According to their figure, in some cases the TR line exhibits a blue shift or a blue wing in the ribbon front profiles (blue curves) although the most likely profiles (red curves) are redshifted. However, the profiles also often exhibit spectral characteristics, including absorption features, which are typical of optically thick conditions (see for example the profiles in the top left multi-panel in their figure). We suggest that a future statistical study using ML on the optically thin \oiv~1401~\AA~instead of the \siiv~lines (for observations where the \oiv~line is visible enough), might provide useful insights into the behaviour of the TR in the ribbon leading edge and thus crucial constraints on the models.

Based on our preliminary results, we conclude that the optically thin TR lines such as \oiv~\textit{can} be blueshifted in the ribbon front profiles, in agreement with our speculation based on the \radynfp~models.  

It is also possible that the regional signatures that characterise the ribbon front in the TR lines are not as a clear as those observed in the chromospheric lines, and that the TR lines there can be both blue and redshifted.

On the other hand, since we can already explain the cases when the \oiv~lines are blueshifted, we investigated whether we can reproduce \textit{redshifted} TR emission for our gradual gentle heating models by adding  \textsl{in-situ} heating in the corona. Previous work \citep[e.g.][]{2014Sci...346B.315T,Polito2018} has in fact shown that \textsl{in-situ} typically produces downflows in the TR. 

To model the effects of \textsl{in-situ} energy deposition and the subsequent conductive heat flux through the corona to the transition region and chromosphere, we experimented with adding an additional \textsl{ad-hoc} volumetric heating rate to the looptop in the $F_{peak} = 5\times10^{9}$ erg~s$^{-1}$~cm$^{-2}$, $\delta = 5$, E$_{c}$ = 30~keV, $t_{inj} = 20$~s electron beam simulation. These volumetric heating rates were $Q_{adhoc} = [1.0, 2.5, 5.0, 7.5, 10]$~erg~s$^{-1}$~cm$^{-3}$, applied for 10s over the top $200$~km of the loop (giving instantaneous energy fluxes of $F_{adhoc} = [0.2, 0.5, 1.0, 1.5, 2.0]$~$\times10^{9}$~erg~s$^{-1}$~cm$^{-2}$).   

We found that (not shown here) these hybrid simulations do not reproduce redshifted TR emission and the typical ribbon front profiles simultaneously.

\section{Conclusions}
\label{Sect:conclusions}

In this work we have analysed the spectral characteristics of the \iris~\mgii~k, \mgii~triplet, \fexxi~and \oiv~lines in flare ribbons for four different flares, particularly focusing on the regional differences between the so-called ribbon ``front" and ``trailing" profiles \citep{Xu2016,Panos2018,Panos2021a,Panos2021b}. We have quantified the spectral characteristics of the typical \mgii~k profiles to allow a detailed comparison with radiative hydrodynamic simulations using the \radynfp~and \rhtiago~codes. We have also studied the correlation between the location of \mgii~ribbon leading edge profiles and the intensity and flows observed in the \fexxi~line (the latter being a signature of chromospheric evaporation) and \oiv~TR line. The key results from our observational and modelling analysis can be summarised as follows:
\begin{itemize}
    \item The location of the ribbon front profiles in the \iris~\mgii~ chromospheric lines does not coincide most often with the location of strongest chromospheric evaporation and TR emission. 
    \item Our \radynfp~ simulations suggest that heating models where the electrons have a more modest and gradual energy flux and higher E$_C$  can qualitatively reproduce the observed enhanced \mgii~central reversals and \mgii~triplet emission, but do not deposit enough energy to drive explosive chromospheric evaporation. 
    \item Models with larger flux and smaller E$_C$ which are better at driving explosive evaporation and heating of the TR, cannot explain the ribbon front profiles. 
    \item The optically thin \oiv~1401\AA~line \textit{can} be blueshifted in the ribbon front profiles, in agreement with our \radynfp~ models shown here.  However, a more extensive analysis of optically thin \oiv~spectra for a larger sample of flares, possibly including ML analysis similar to that in the Panos et al. papers, would be needed to confirm our preliminary results.
\end{itemize}

One thing our models still cannot fully explain is the long duration (up to a few minutes) of the \mgii~ribbon front profiles \citep[e.g.][]{Panos2018} and enhanced \hei~line absorption \citep{Xu2016,2021ApJ...912..153K}. We are currently focused on addressing this issue as part of a follow up paper (Kerr et al, \textsl{In Prep.}).

Based on the results above, we propose that different heating processes might be at play in different regions of the ribbons at a certain time: 
\begin{enumerate}
    \item harder electron spectra but with a more gentle flux might be responsible for the enhanced reversal of the \mgii\ lines \citep[in agreement with][]{2021ApJ...912..153K};
    \item different populations of accelerated electrons with higher fluxes \& lower E$_C$ drive chromospheric evaporation and the heating of the upper atmosphere.
\end{enumerate} 

It is not clear if an evolution of the electron beam heating parameters over time might be able to explain both behaviours in a consistent way. We are aiming to address this question in Paper II (Kerr et al, \textsl{In Prep.}).

Finally, our work shows that by combining high-resolution observations by the \iris~spectrograph with advanced hydrodynamic simulations we can obtain crucial constraints on the flare heating models. Future \iris~observations including recent new very high (sub-second) cadence datasets\footnote{https://docs.google.com/document/d/1iPQPTYPULzrnjbnN38j6j0p2AccPTJc1xbqEXZmr0mU/edit} might soon reveal even more interesting and puzzling features in the evolution and characteristics of the ribbon front profiles.

\vspace{1cm}

\textsc{Acknowledgments:} \small{VP acknowledges financial support from the NASA ROSES Heliophysics Guest Investigator program (Grant\# NASA 80NSSC20K0716). GSK acknowledges financial support from NASA's Early Career Investigator Program (Grant\# NASA 80NSSC21K0460). VP, GSK, and YX acknowledge financial support from the NASA ROSES Heliophysics Supporting Research program (Grant\# NASA 80NSSC19K0859). VP And JL acknowledge support from NASA under contract NNG09FA40C ({\it IRIS}). VMS acknowledges the NSF FDSS grant 1936361. This manuscript benefited from discussions held at meetings of the International Space Science Institute team: ``Interrogating Field-Aligned Solar Flare Models: Comparing, Contrasting and Improving,'' led by Dr.~G.~S.~Kerr and Dr.~V.~Polito. IRIS is a NASA small explorer mission developed and operated by LMSAL with mission operations executed at NASA Ames Research center and major contributions to downlink communications funded by the Norwegian Space Center (NSC, Norway) through an ESA PRODEX contract. CHIANTI is a collaborative project involving George Mason University, the University of Michigan (USA), University of Cambridge (UK) and NASA Goddard Space Flight Center (USA).Resources supporting this work were provided by the NASA High-End Computing (HEC) Program through the NASA Advanced Supercomputing (NAS) Division at Ames Research Center.
}

\bibliographystyle{aasjournal}
\bibliography{Polito_etal_ribbonfronts}

\appendix

\section{Additional plots}
\begin{figure*}[!hb]
\center
\includegraphics[width=\textwidth]{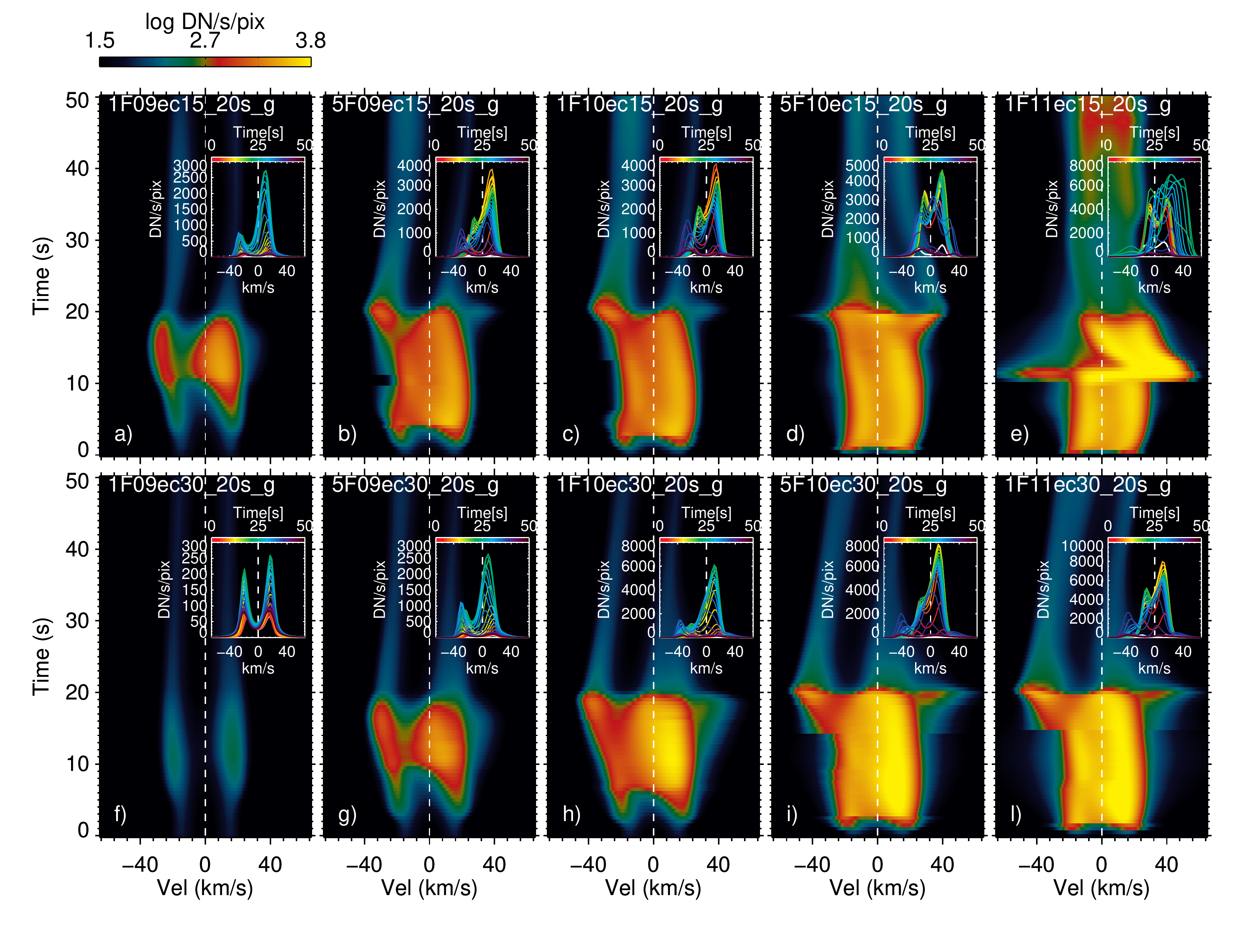}
\caption{\mgii~k spectra for gradual heating models. } 
\label{Fig:mgii_gradual}
\end{figure*}

\begin{figure*}[!hb]
\center
\includegraphics[width=\textwidth]{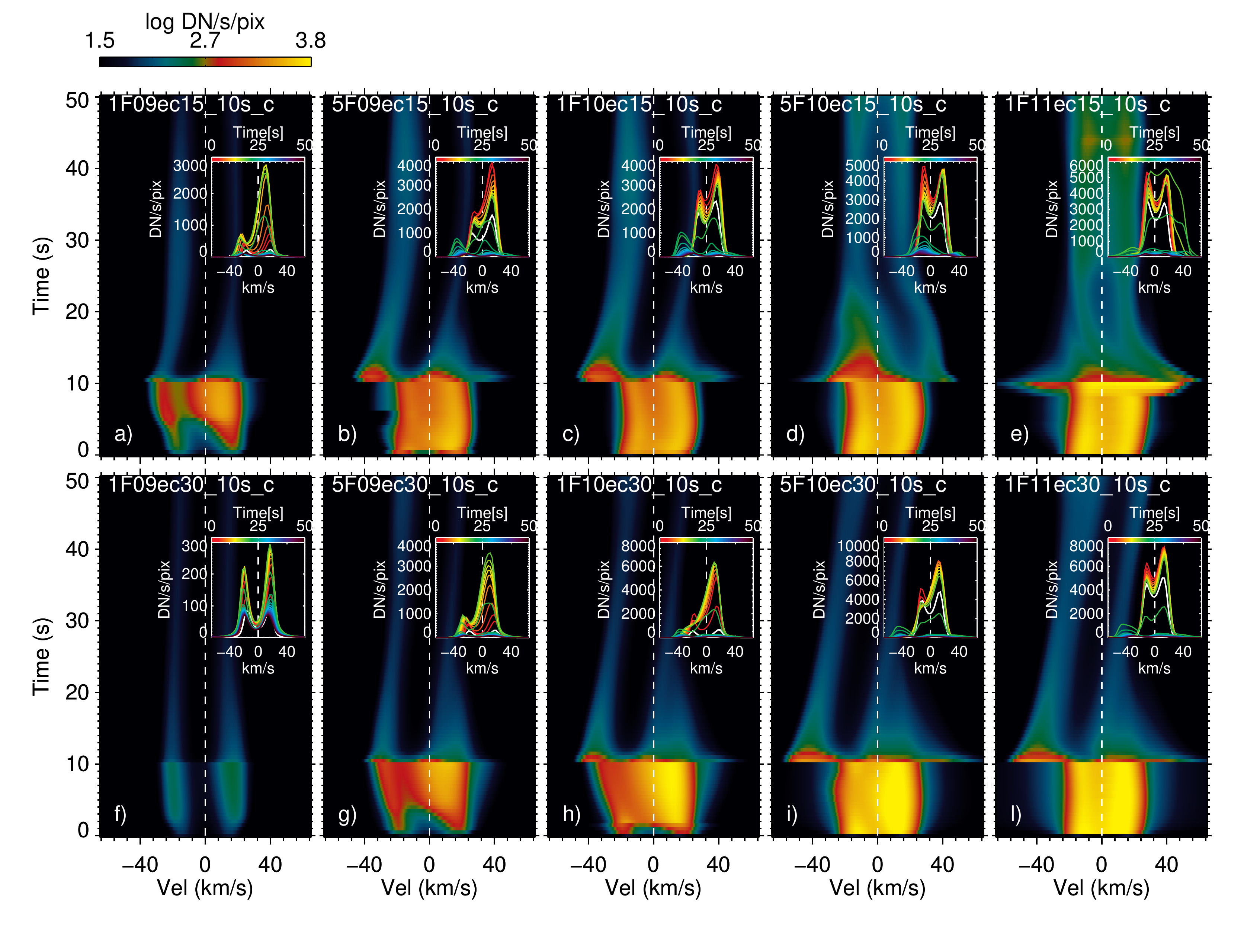}
\caption{\mgii~k spectra for constant heating models.} 
\label{Fig:mgii_const}
\end{figure*}

\begin{figure*}[!hb]
\center
\includegraphics[width=\textwidth]{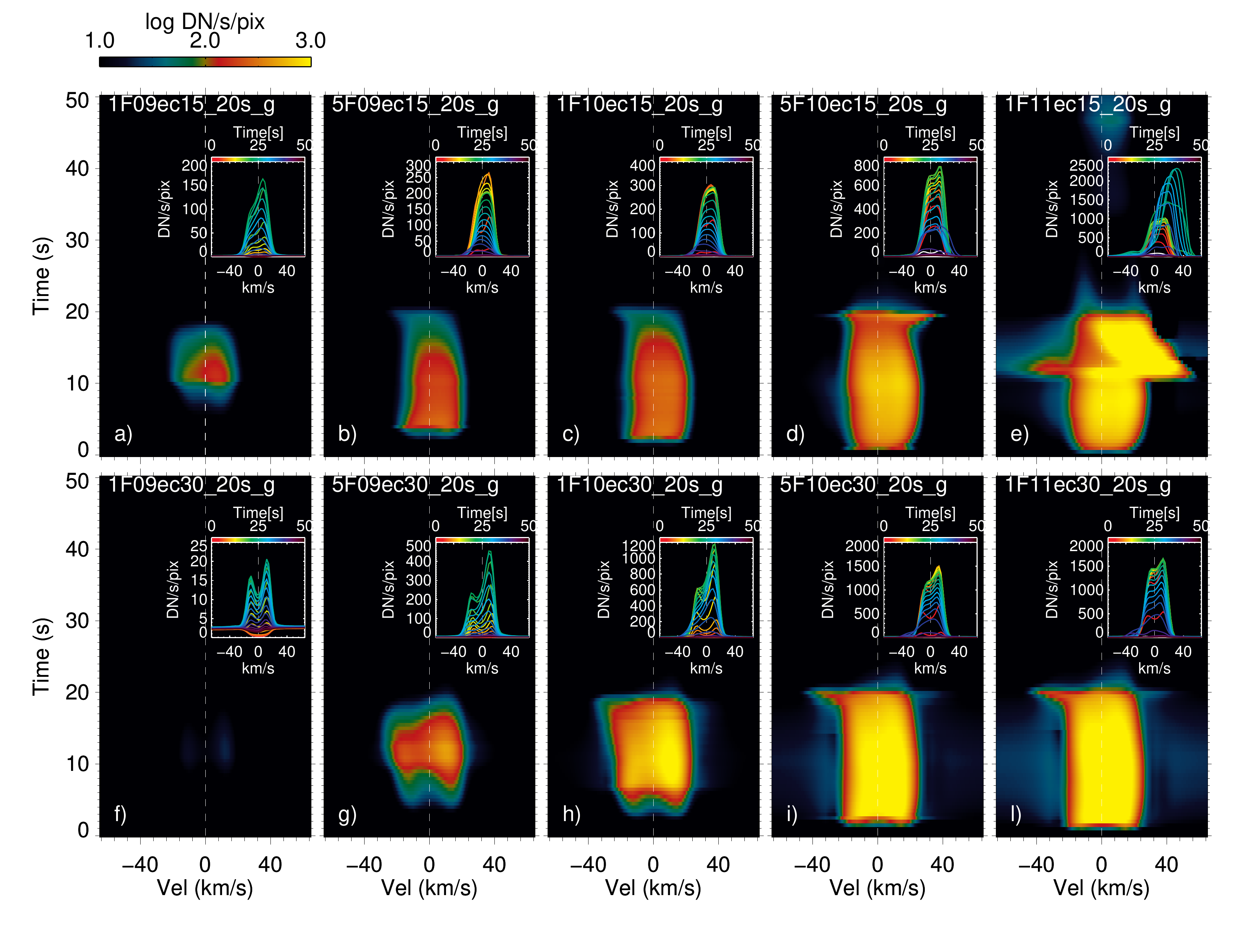}
\caption{\mgii~triplet spectra for gradual heating models.} 
\label{ig:mgiitripl_gradual}
\end{figure*}

\begin{figure*}[!hb]
\center
\includegraphics[width=\textwidth]{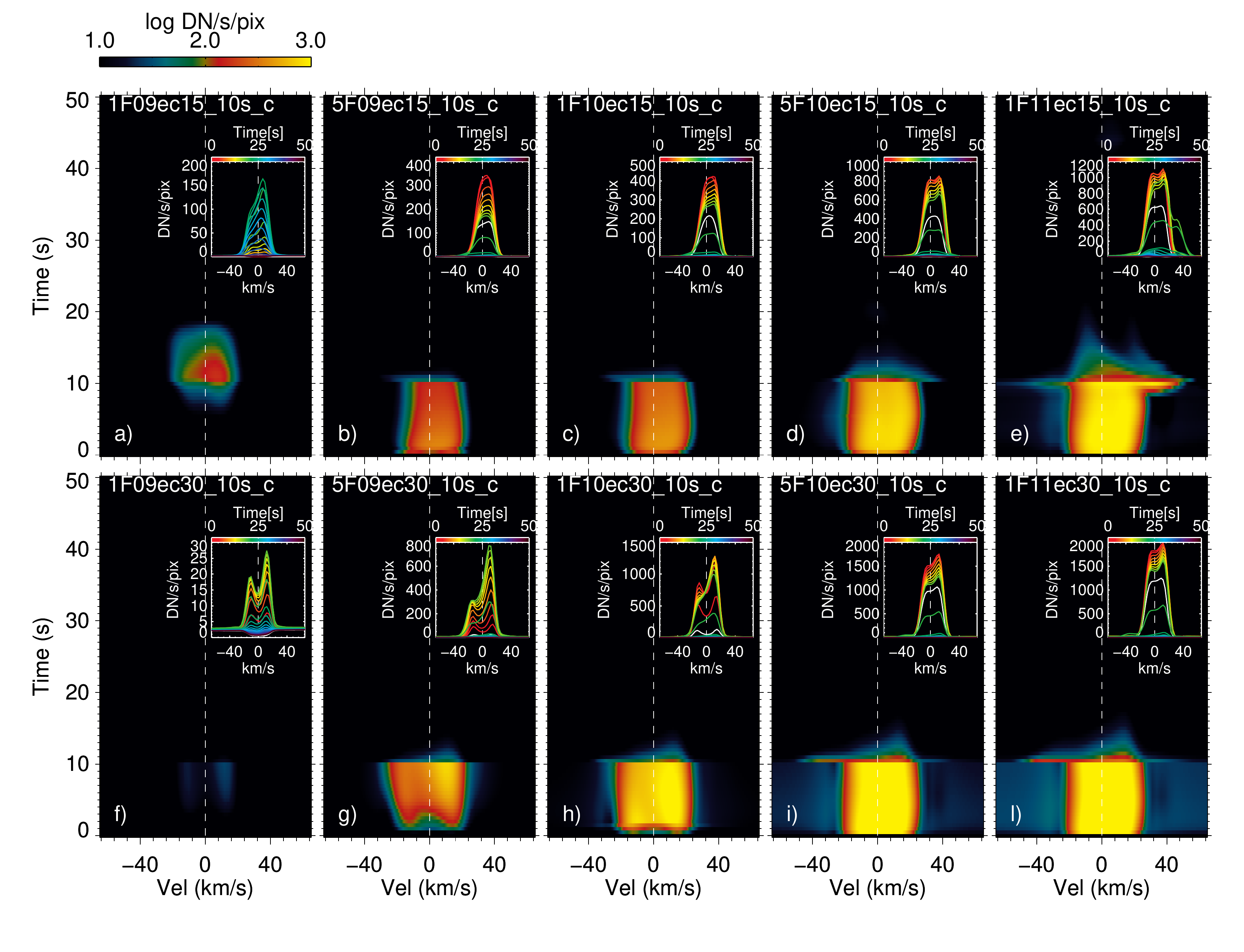}
\caption{\mgii~triplet spectra for constant heating models} 
\label{Fig:mgii_tripl_const}
\end{figure*}

In Figures~\ref{Fig:mgii_gradual}--\ref{Fig:mgii_tripl_const} we show the synthetic spectra of \mgii~k and \mgii~triplet for all \radynfp+\rhtiago~models. For the descriptions of the individual panels, see Figure~\ref{Fig:summary_sims} and text.

\begin{figure*}[!hb]
\center
\includegraphics[width=\textwidth]{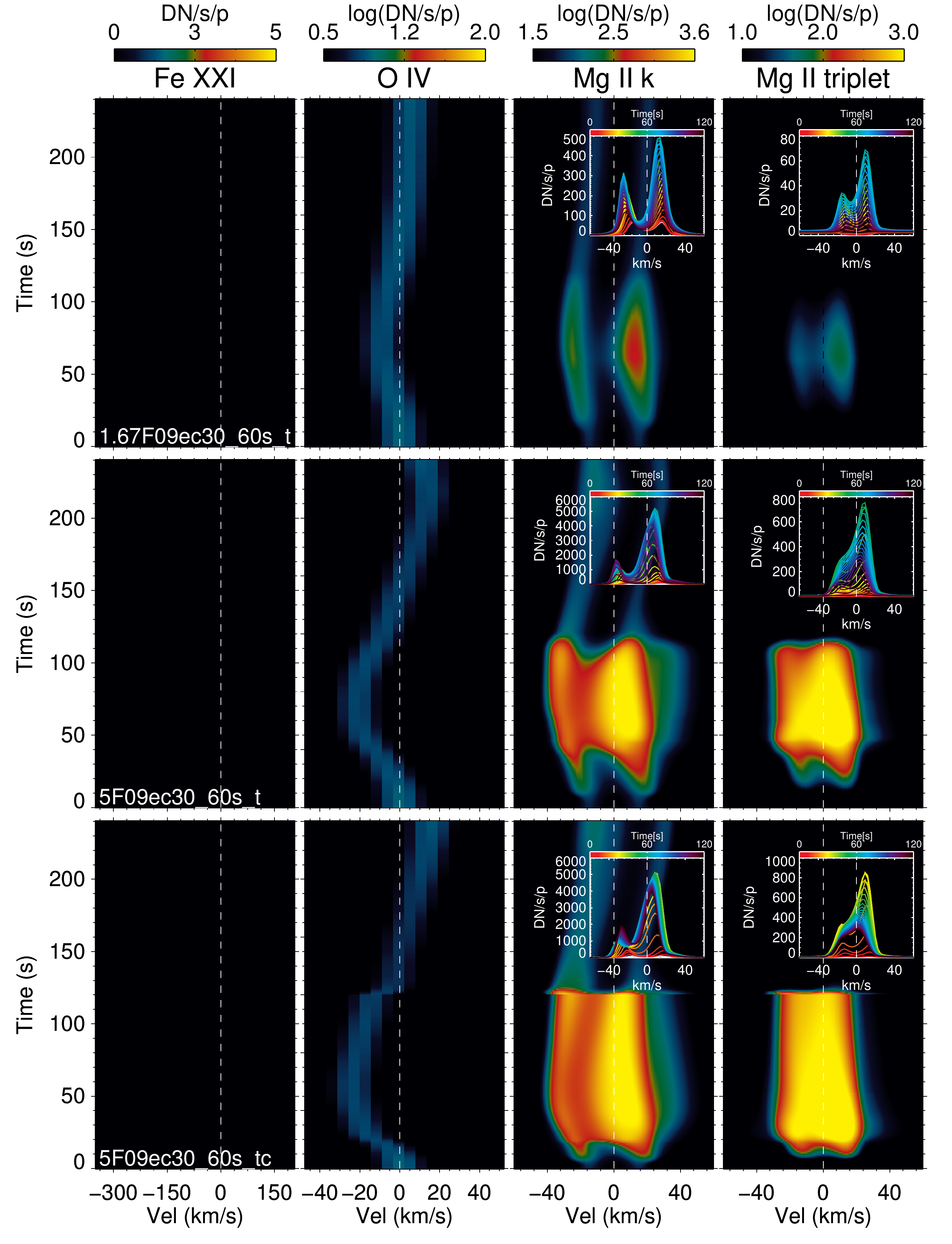}
\caption{Long duration simulations.} 
\label{Fig:long_sims}
\end{figure*}

Figure~\ref{Fig:long_sims} shows the synthethic spectra for the long duration test simulations (see Sect.~\ref{Sect:extra}). 

\begin{figure}
and that \center\offinterlineskip
\includegraphics[width=0.3\textwidth]{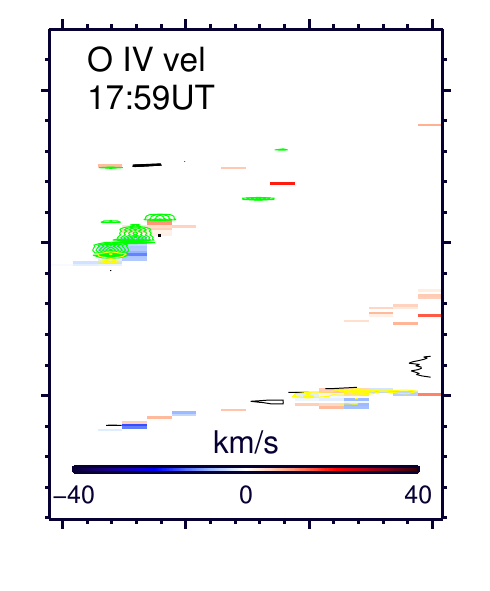}
\includegraphics[width=0.3\textwidth]{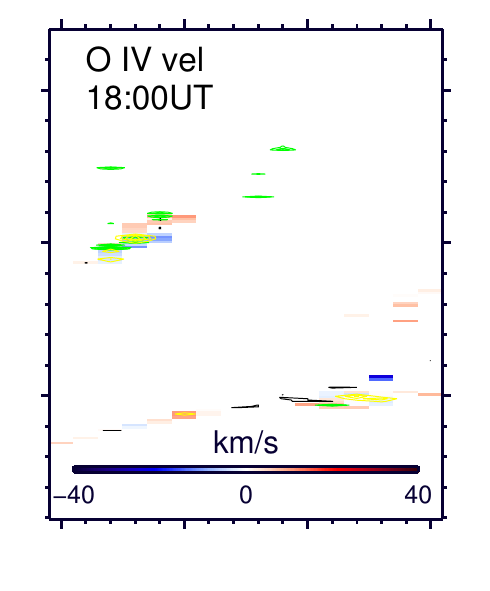}
\caption{Example of \oiv~Doppler Shift maps for two rasters during the 2015-06-22 flare (FL2). The green, yellow and black contours show the location of: the ribbon leading edge pixels, the maximum \oiv~intensity and the maximum \fexxi~evaporation. See text for more details. } 
\label{Fig:doppler2}
\end{figure}

Finally, Figure~\ref{Fig:doppler2} shows the \oiv~Doppler shift maps for two rasters during the FL2 (see Sect.~\ref{Sect:TR}).


\end{document}